\documentclass[aps,prl,twocolumn,superscriptaddress,floatfix,noeprint,10pt]{revtex4-2}

\usepackage{graphicx,makecell}
\usepackage{amsmath,amsfonts,amssymb}
\usepackage{mathtools}
\usepackage{mathrsfs}
\usepackage{bm,dsfont}
\usepackage[dvipsnames]{xcolor}
\usepackage[colorlinks=true, citecolor=Green, linkcolor=BrickRed, urlcolor=NavyBlue]{hyperref}
\usepackage[all]{hypcap}

\begin{document}

\title{
Extrinsic quantum geometry in the quadrupolar bulk photovoltaic effect
}

\author{Steven Gassner}
\affiliation{Department of Physics and Astronomy, University of Pennsylvania, Philadelphia, Pennsylvania 19104, USA}

\author{Swati Chaudhary}
\affiliation{The Institute for Solid State Physics, The University of Tokyo, Kashiwa, Chiba 277-8581, Japan
}

\author{Martin Claassen}
\affiliation{Department of Physics and Astronomy, University of Pennsylvania, Philadelphia, Pennsylvania 19104, USA}

\date{\today}

\begin{abstract}

The bulk photovoltaic effect has become a valuable probe of the quantum geometry of Bloch bands. While it is restricted to inversion-broken systems within the dipole approximation, the finite momentum of light is appreciated to give rise to this effect even in centrosymmetric crystals, an effect referred to as “photon drag.” In this work, we recast the photon drag effect at leading order in the optical wavevector, highlighting a previously neglected contribution arising intuitively from the electric quadrupole correction to light-matter coupling. In the language of band geometry, we identify this interband quadrupole as a multiband metric tensor that quantifies the variation of two resonantly driven states extrinsic to the subspace they span. We predict that systems in which three or more bands strongly admix in momentum space, such as twisted MoTe$_2$ bilayers, will have anomalously large photon drag due to this quadrupolar effect. Our work provides a conceptual bridge between band-geometric organizing principles and electromagnetic multipole corrections in nonlinear optics.

\end{abstract}

\maketitle

The detailed understanding of the interaction between light and electrons in solids is one of the most celebrated successes of quantum theory. In large part, the optical properties of atoms and materials can be understood in terms of the absorption and emission of photons by electrons undergoing virtual transitions between energy eigenstates \cite{sipe1993,boyd2008}. A key consequence is that effective photon-photon interactions can arise in materials with strong optical matrix elements between their electronic states, enabling highly nonlinear optical phenomena. Modern treatments of these optical responses highlight the central role of band geometry: interpreting the electric dipole matrix element between quantum states in a band structure as a tangent vector defines a manifold of quantum states parameterized by momentum whose geometry largely determines the magnitude of the response \cite{ahn2020, ahn2022,dai2023recent,morimoto2016topological,nagaosa2017concept,nagaosaNonlinearOpticalResponses2022,OrensteinJ2021TaSo,parker2019}. This variation of states in momentum space introduces a ``geometric length scale," which can be many times the lattice constant \cite{verma2026} and is especially pronounced and tunable in recently studied twisted two-dimensional materials with flat Chern bands \cite{bistritzer2011,cao2018,xu2023,kennes2021}.
\begin{figure}[t]
    \centering
    \includegraphics[width=0.95\linewidth]{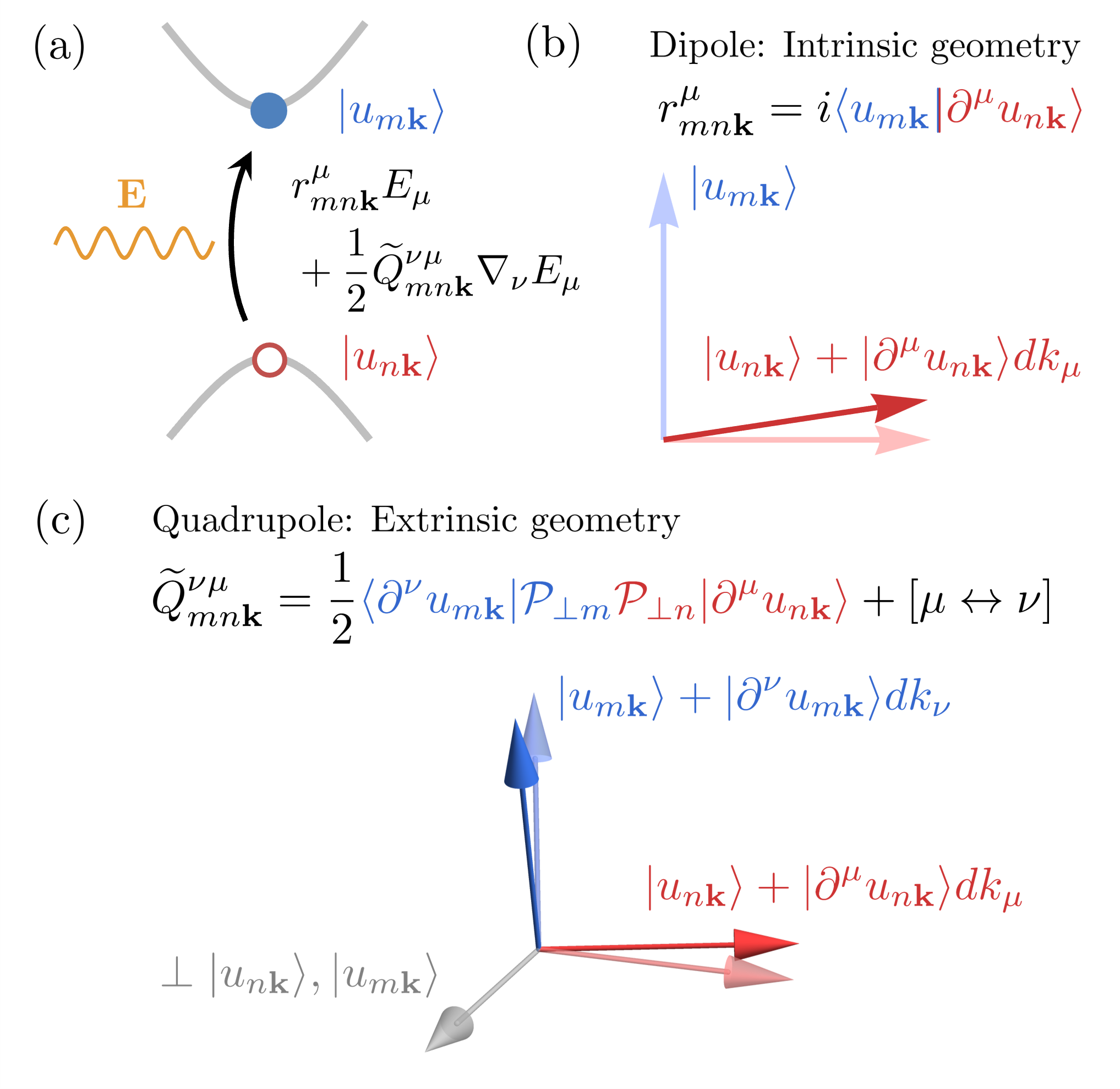}
    \caption{
    Electric quadrupole corrections to optical responses encode extrinsic quantum geometry.
    (a) Electrons in Bloch bands couple to a uniform electric field via an interband electric dipole matrix element. Gradients in the electric field couple to an interband quadrupole matrix element.
    (b) The interband electric dipole $r^\mu_{mn\mathbf{k}}$ probes the \textit{intrinsic} geometry of the Bloch states, i.e. within the resonant two-band manifold. (c) The origin-independent part of the interband electric quadrupole, which we denote as $\widetilde{Q}^{\nu\mu}_{mn\mathbf{k}}$, probes the \textit{extrinsic} geometry of the Bloch states, i.e. orthogonal to the resonant two-band manifold. Here $\mathcal{P}_{\perp n} = 1-|u_{n\mathbf{k}}\rangle\langle u_{n\mathbf{k}}|$ is the projector to the subspace orthogonal to $|u_{n\mathbf{k}}\rangle$.}
    \label{fig:extrinsic geometry}
\end{figure}

It is well-known, however, that treating optical responses in terms of the electric dipole alone is an approximation that neglects dependence on the wavevector of light~\cite{ocana2023,malashevich2010,mckay2024}. Expanding responses in powers of the wavevector is standard and can be understood as a multipole expansion: for instance, the linear-in-$\mathbf{q}$ corrections are conventionally separated into electric quadrupole and magnetic dipole parts. What justifies zeroth-order truncation in most cases is that the wavelength of a photon resonant with an electronic excitation is usually many times larger than the associated dipole moment. Even so, especially in cases where symmetry forbids the zeroth-order contribution, essential features of certain optical responses require an understanding beyond the dipole approximation.

It is a largely open question how beyond-dipole corrections enter the band-geometric framework for nonlinear optical responses in solids. A prime target for investigating this question is the bulk photovoltaic effect (BPVE), a second-order response generating a static current from finite-frequency light in a homogeneous material. We focus on the BPVE for two reasons: (i) the so-called ``shift current" contribution is well understood as a purely quantum geometric response~ \cite{ahn2022,ahn2020,holder2020,dai2023recent,nagaosa2017concept,morimoto2016topological,OrensteinJ2021TaSo}, and (ii) the response is forbidden by inversion symmetry within the dipole approximation. Finite-wavevector photocurrents have been extensively studied under the name ``photon-drag effects" \cite{ribakovs1977,gibson1977}, with recent work highlighting the geometric aspects of the response \cite{shi2021,xie2025,lu2025,Qu2025}. In this work, we reformulate the linear-in-$\mathbf{q}$ corrections in terms of multipole moments \cite{ocana2023,gassner2023,malashevich2010}, highlighting a previously overlooked ``three-band" contribution that can in some cases dominate the response. The crucial observation, summarized in Fig. \ref{fig:extrinsic geometry}, is that a natural definition for the quadrupolar contribution to the BPVE probes the geometry of two resonantly driven states ``\textit{extrinsic}" to the subspace they span \cite{avdoshkin2023}. That is, while the extensively studied electric dipole responses are limited to geometric quantities ``intrinsic" to the resonant two-band subspace, the identification of the interband quadrupole moment as a \textit{non-Abelian} metric tensor captures additional extrinsic geometric contributions that, as we show, can importantly modify the response.

\paragraph{Band geometry of multipolar optical responses} Consider a light-matter-coupled single-particle Hamiltonian describing electrons with charge $-e<0$ in velocity gauge $\mathcal{H}_\mathbf{A} = \mathcal{H}_0 + e\sum_{\mathbf{q}_j} j^\alpha_{\mathbf{q}_j} A_{\alpha}(\omega_j,\mathbf{q}_j) + \mathcal{O}(\mathbf{A}^2)$, expressed in terms of a charge-normalized current operator $j^\alpha_\mathbf{q}$ and vector potential whose time derivative encodes the electric field $\mathbf{A}(\omega,\mathbf{q}) \equiv \mathbf{E}(\omega,\mathbf{q})/i\omega$. Here, $\mathcal{H}_0$ describes a band insulator with Bloch states $\mathcal{H}_0|\psi_{n\mathbf{k}}\rangle = \hbar\omega_{n\mathbf{k}}|\psi_{n\mathbf{k}}\rangle$ with $|\psi_{n\mathbf{k}}\rangle = e^{i\mathbf{k}\cdot\mathbf{r}}|u_{n\mathbf{k}}\rangle$. To understand the wavelength-dependent BPVE, a natural starting point is to inspect the first-order-in-$\mathbf{q}$ correction to the current operator matrix element for interband ($m\neq n$) transitions
\begin{align}
    \frac{\langle\psi_{m,\mathbf{k}_+}|j^\mu_{\mathbf{q}}|\psi_{n,\mathbf{k}_-}\rangle}{i\left(\omega_{m,\mathbf{k}_+}-\omega_{n,\mathbf{k}_-}\right)} &\approx r^\mu_{mn\mathbf{k}} + q_\nu \bigg(\frac{i}{2}\widetilde{Q}^{\nu\mu}_{mn\mathbf{k}} + \frac{ \varepsilon^{\nu\mu} \widetilde{M}_{mn\mathbf{k}} }{\omega_{mn\mathbf{k}}} \notag\\
    &\hspace{1cm}+\frac{\mathcal{A}^\nu_{m\mathbf{k}}+\mathcal{A}^\nu_{n\mathbf{k}}}{2}r^\mu_{mn\mathbf{k}}\bigg)
    \label{multipoles}
\end{align}
where $\mathbf{k}_\pm \equiv \mathbf{k} \pm \frac{\mathbf{q}}{2}$, $r_{mn\mathbf{k}} = \langle u_{m,\mathbf{k}}|\partial^\mu u_{n,\mathbf{k}}\rangle$ is the conventional $q=0$ interband dipole matrix element, $\partial^\mu$ denotes a momentum derivative, and repeated indices ($\nu$) are implicitly summed. In this expression, we have made two key separations: First, we identify the symmetric and antisymmetric $\sim \mathbf{q}$ corrections with respect to $[\mu\leftrightarrow\nu]$ as the electric quadrupole and magnetic dipole matrix elements, respectively. Second, as is discussed in Ref.~\cite{ocana2023}, we explicitly separate out contributions that depend on the intraband Berry connection $\mathcal{A}^\nu_{n\mathbf{k}} \equiv i \langle u_{n\mathbf{k}}|\partial^\nu u_{n\mathbf{k}}\rangle$, which we combine in the final term of Eq.~(\ref{multipoles}). This defines the manifestly \textit{origin-independent} electric quadrupole $\widetilde{Q}^{\nu\mu}_{mn\mathbf{k}}$ and  magnetic dipole $\widetilde{M}_{mn\mathbf{k}}$~\cite{ocana2023}, which are invariant under shifts of origin within the unit cell (which generally changes $\mathcal{A}^\nu_{n\mathbf{k}}$). We denote the multipole moments with a tilde to emphasize the distinction from their traditional origin-dependent definitions. 
The magnetic dipole coupling is often defined with a spatial direction $B_\rho(\omega)M^\rho_{mn\mathbf{k}}$, which we suppress here, implicitly fixed via Faraday's law $B^\rho(\omega,\mathbf{q}) = q_\nu\frac{\varepsilon^{\nu\mu\rho}}{\omega}E_\mu(\omega,\mathbf{q})$. We now express these moments in terms of $\mathbf{k}$-derivatives of the cell-periodic Bloch states $|u_{n\mathbf{k}}\rangle$ and $|u_{m\mathbf{k}}\rangle$ as
\begin{align}
    \widetilde{Q}^{\mu\nu}_{mn\mathbf{k}} &= \frac{1}{2}\langle \partial^\mu u_{m\mathbf{k}} | \mathcal{P}^\perp_{ m\mathbf{k}}\mathcal{P}^\perp_{n\mathbf{k}} | \partial^\nu u_{n\mathbf{k}}\rangle + [\mu\leftrightarrow\nu] ,
    \label{quadrupole moment} \\\varepsilon^{\mu\nu}\widetilde{M}_{mn\mathbf{k}} &= \frac{1}{2}\langle \partial^\nu u_{m\mathbf{k}}|\mathcal{P}^\perp_{m\mathbf{k}} v^\mu_{\mathbf{k}} | u_{n\mathbf{k}}\rangle  \notag\\
    &+ \frac{1}{2} \langle u_{m\mathbf{k}}| v^\mu_{\mathbf{k}} \mathcal{P}^\perp_{n\mathbf{k}} | \partial^\nu u_{n\mathbf{k}}\rangle - [\mu\leftrightarrow\nu]
\end{align}
where $\mathcal{P}^\perp_{n\mathbf{k}} \equiv 1-|u_{n\mathbf{k}}\rangle\langle u_{n\mathbf{k}}|$.

A crucial feature of these interband multipole moments that is illustrated by their band-geometric interpretation is that they quantify the $\mathbf{k}$-space variation of the initial and final states in the directions \textit{orthogonal to both states} in Hilbert space. The quadrupole moment in particular is purely a non-Abelian metric tensor, generalizing the standard Fubini-Study metric to the case of a \textit{pair} of bands that are placed in resonance via the external field. We refer to geometry quantified by this metric as ``extrinsic" in the sense that it is extrinsic to the two-state space $\{|u_{n\mathbf{k}}\rangle, |u_{m\mathbf{k}}\rangle\}$ being resonantly driven. This implies that origin-independent multipolar effects are fundamentally \textit{multiband} in that they will be missed in effective models of less than three total bands. We note that similar multiband effects have been noted in anomalous Hall effects at nonlinear order in the electric field \cite{kaplan2023}.

\paragraph{Quadrupolar bulk photovoltaic effect} While the BPVE is forbidden in centrosymmetric materials within the dipole approximation, multipolar light-matter interactions can enable a finite photocurrent that is proportional to the wavelength of the incident field:
\begin{equation}
    J^\mu = \sigma_{(1),\textrm{PV}}^{\nu\mu\alpha\beta}(\omega)~ q_\nu~ E_\alpha(\omega,\mathbf{q}) E_{\beta}(-\omega,-\mathbf{q})  \label{eq:shift quadrupole}
\end{equation}
Here, $\mu$ is the direction of the induced DC photocurrent, $\alpha,\beta$ are polarization indices of the applied electric field $E_\alpha(\omega,\mathbf{q}) = \int d\omega\int d^d\mathbf{q}\, e^{-i(\mathbf{q}\cdot\mathbf{r}-\omega t)} E_\alpha(t,\mathbf{r})$, and $\nu$ is set by the field's angle of incidence.

To distill the quantum-geometric multipole origin of this response, we start by computing the $\mathbf{q}$-dependent response $\langle j^\mu \rangle(0,\mathbf{0}) = \sigma^{\mu\alpha\beta}_\text{PV}(\omega,\mathbf{q}) E_\alpha(\omega,\mathbf{q})E_\beta(-\omega,-\mathbf{q})$ and expand $\sigma^{\mu\alpha\beta}_\text{PV}(\omega,\mathbf{q})$ to linear order in $\mathbf{q}$, justified by the disparity of photon and lattice wavelengths for optical or THz electric fields. We obtain 
\begin{align}
    \sigma^{\mu\alpha\beta}_\text{PV}(\omega,\mathbf{q}) = &-\frac{\pi e^3}{\hbar^2\omega^2}\int_\mathbf{k}\sum_{mn} f_{m\mathbf{k}_+,n\mathbf{k}_-}\delta(\omega_{m\mathbf{k}_+,n\mathbf{k}_-}-\omega) \notag\\
    &~~~~~~~~~\times \left(\mathcal{I}^{\mu\alpha\beta,(\text{shift})}_{m\mathbf{k}_+,n\mathbf{k}_-} + \mathcal{I}^{\mu\alpha\beta,(\text{inj.})}_{m\mathbf{k}_+,n\mathbf{k}_-}\right)
    \label{sigma PV q}
\end{align}
with shift and injection current matrix elements
\begin{align}
\mathcal{I}^{\mu\alpha\beta,(\text{shift})}_{m\mathbf{k}_-,n\mathbf{k}_+} &= j^\beta_{n\mathbf{k}_-,m\mathbf{k}_+}\left(i\nabla^\mu_{\mathbf{k}}+\mathcal{A}^\mu_{m\mathbf{k}_+}-\mathcal{A}^\mu_{n\mathbf{k}_-}\right)j^\alpha_{m\mathbf{k}_+,n\mathbf{k}_-} \notag\\
    &+ [\alpha\leftrightarrow\beta]^*
,\label{matrix element shift kprime k} \\
    \mathcal{I}^{\mu\alpha\beta,(\text{inj})}_{m\mathbf{k}_-,n\mathbf{k}_+} &=\tau (v^\mu_{m\mathbf{k}_+}-v^\mu_{n\mathbf{k}_-})j^\beta_{n\mathbf{k}_-,m\mathbf{k}_+}j^\alpha_{m\mathbf{k}_+,n\mathbf{k}_-} \notag\\
    &+ [\alpha\leftrightarrow\beta]^*
    \label{matrix element injection kprime k}.
\end{align}
Here, $\int_\mathbf{k} \equiv \int_\text{BZ}d^d\mathbf{k}/(2\pi)^d$ denotes a Brillouin zone average in $d$ dimensions,  $\omega_{m\mathbf{k}',n\mathbf{k}}\equiv \omega_{m\mathbf{k}'}-\omega_{n\mathbf{k}}$ encodes the energy difference between bands $n$, $m$;  $f_{m\mathbf{k}_+,n\mathbf{k}_-} \equiv f(\omega_{m\mathbf{k}_+}) - f(\omega_{n\mathbf{k}_-})$ with Fermi functions $f(\cdot)$, $\mathcal{A}^\mu_{n\mathbf{k}} \equiv i\langle u_{n\mathbf{k}}|\partial_{k_\mu}u_{n\mathbf{k}}\rangle$ is the intraband Berry connection, $v^\mu_{n\mathbf{k}}\equiv \partial_{k_\mu}\omega_{n\mathbf{k}}$ is the intraband velocity, $j^\alpha_{m\mathbf{k}',n\mathbf{k}} = \langle \psi_{m\mathbf{k}'}|j^\alpha_{\mathbf{k}'-\mathbf{k}}|\psi_{n\mathbf{k}}\rangle$ is the finite-wavector current operator in Bloch representation, and $[\alpha\leftrightarrow\beta]^*$ denotes the complex conjugate after interchange of indices $\alpha$ and $\beta$. Here we are careful to define the response in terms of the current operator matrix elements, which need not coincide with the often-used approximation $j^\alpha_{m\mathbf{k}'} \approx \langle u_{m\mathbf{k}'} | v^\alpha_{\frac{\mathbf{k}'+\mathbf{k}}{2}} | u_{n\mathbf{k}}\rangle$ in the photon-drag literature (see Appendix A). Eqs. (\ref{matrix element shift kprime k}) and (\ref{matrix element injection kprime k}) correspond to the shift and injection current, respectively. The latter is controlled by a phenomenological scattering time $\tau$. In what follows, we focus our attention on the shift current contribution.

The quadrupolar BPVE $\sigma_{(1),\textrm{PV}}^{\nu\mu\alpha\beta}(\omega)$ [Eq. (\ref{eq:shift quadrupole})] can now be computed via expanding to linear order in $\mathbf{q}$, resulting in two contributions: First, an ``energetic resonance'' term
\begin{align}
\sigma^{\nu\mu\alpha\beta}_{(1),\text{res.}}(\omega) &= -\tfrac{\pi e^3}{\hbar^2}\int_\mathbf{k}\sum_{mn}f_{mn\mathbf{k}}\delta'(\omega_{mn\mathbf{k}}-\omega)\tfrac{v^\nu_{m\mathbf{k}}+v^\nu_{n\mathbf{k}}}{2}\,\mathcal{I}^{\mu\alpha\beta}_{mn\mathbf{k}}
    \label{sigma ene}
\end{align}
which depends on the intraband velocity and vanishes for flat bands. Here, $\mathcal{I}^{\mu\alpha\beta}_{mn\mathbf{k}} \equiv \lim_{\mathbf{q}\to 0} \mathcal{I}^{\mu\alpha\beta}_{m,\mathbf{k}+\frac{\mathbf{q}}{2};n,\mathbf{k}-\frac{\mathbf{q}}{2}}$ is the integrand of the ordinary BPVE. Second, we find a multipolar contribution
\begin{equation}
\begin{split}
    \sigma^{\nu\mu\alpha\beta}_{(1),\text{mul.}}(\omega) = -\tfrac{\pi e^3}{\hbar^2} \int_\mathbf{k}\sum_{mn}f_{mn\mathbf{k}}\delta(\omega_{mn\mathbf{k}}-\omega)\,\mathcal{I}^{\nu\mu\alpha\beta}_{mn\mathbf{k}}
    ,
    \label{sigma mul}
\end{split}
\end{equation}
where the four-index integrand $\mathcal{I}^{\nu\mu\alpha\beta}_{mn\mathbf{k}}$ is obtained upon substituting Eq. (\ref{multipoles}) into Eq. (\ref{matrix element shift kprime k}):
\begin{align}
    \mathcal{I}^{\nu\mu\alpha\beta}_{mn\mathbf{k}} &= \frac{1}{2}\left(\widetilde{Q}^{\nu\beta}_{nm\mathbf{k}}r^{\alpha;\mu}_{mn\mathbf{k}} -r^\beta_{nm\mathbf{k}}\widetilde{Q}^{\nu\alpha;\mu}_{mn\mathbf{k}}\right) \notag\\
    &- i \left(\varepsilon^{\nu\beta} \frac{\widetilde{M}_{nm\mathbf{k}}}{\omega_{nm\mathbf{k}}} r^{\alpha;\mu}_{mn\mathbf{k}} - \varepsilon^{\nu\alpha} r^\beta_{nm\mathbf{k}}\left(\frac{\widetilde{M}_{mn\mathbf{k}}}{\omega_{mn\mathbf{k}}}\right)^{;\mu}\right) \notag\\
    &+ \frac{1}{2}r^\beta_{nm\mathbf{k}}r^\alpha_{mn\mathbf{k}}\left(\Omega^{\nu\mu}_{m\mathbf{k}} + \Omega^{\nu\mu}_{n\mathbf{k}}\right) ~+~ [\alpha\leftrightarrow\beta]^*.
\end{align}
Here, the first, second, and third line describe the electric quadrupole, magnetic dipole, and anomalous velocity contributions to the BPVE, respectively. Furthermore, $(.)^{;\mu}_{mn} = \partial_{k_\mu}(.)_{mn}-i(\mathcal{A}^\mu_{mm}-\mathcal{A}^\mu_{nn})(.)_{mn}$ denotes the $U(1)^2$ gauge-covariant derivative that appears also in the standard shift current~\cite{sipe2000,ahn2022}. Importantly, for $\nu = \alpha = \beta$, the magnetic dipole contribution vanishes, leaving only electric quadrupole and anomalous velocity contributions.

Time reversal symmetry (TRS) imposes finite-$\mathbf{q}$ selection rules that are analogous to those for the conventional shift and injection currents \cite{lu2025}. In TRS materials, the quadrupolar shift (injection) current vanishes for linear (circular) polarization.

\begin{figure}[t]
    \centering
    \includegraphics[width=0.9\linewidth]{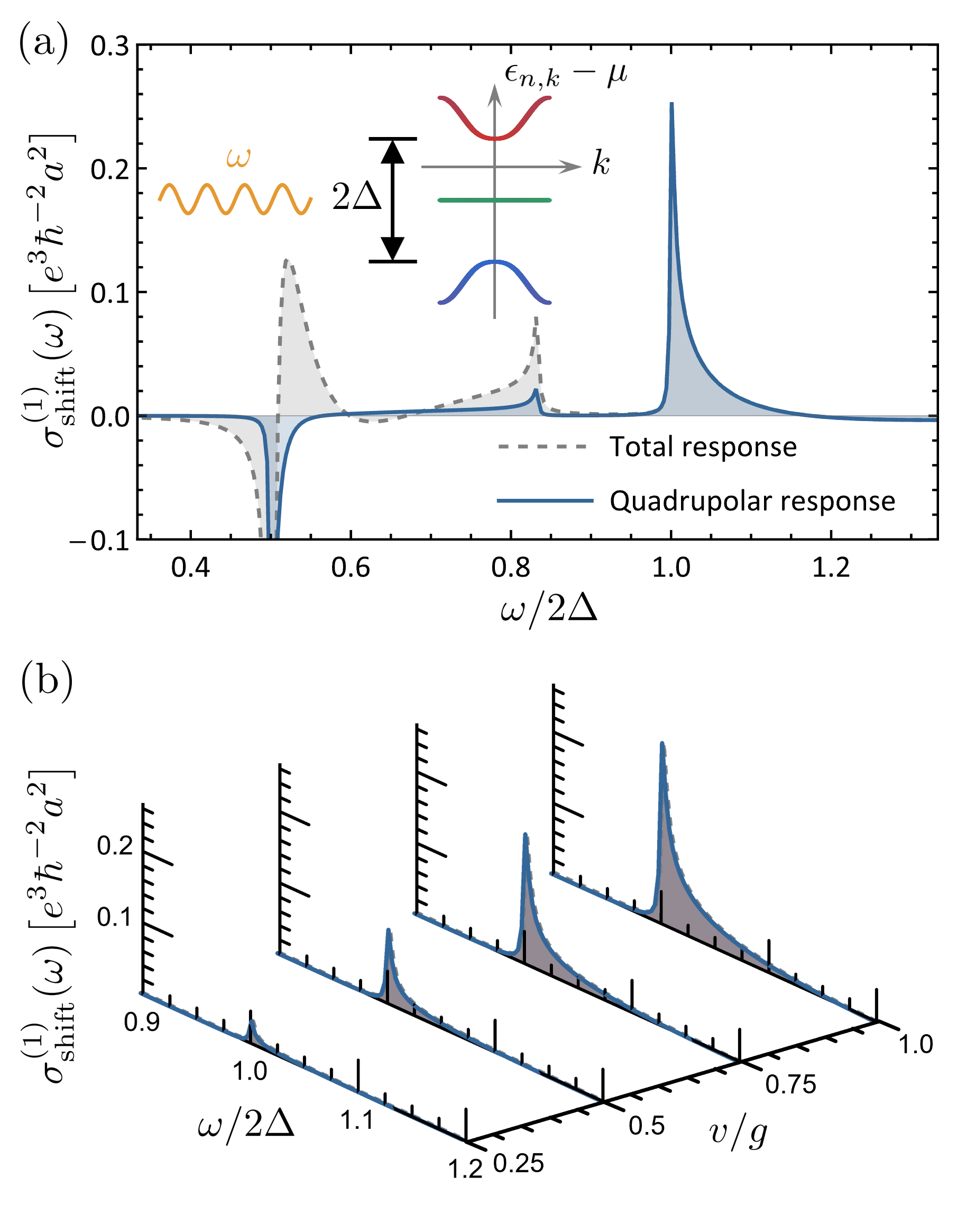}
    \caption{Shift current at linear order in wavevector for the minimal three-band model described in the text, in which near $k=0$ the upper and lower dispersing bands have a mutual coupling $\sim -ig k^2$ and a coupling $\sim -\frac{i}{\sqrt{2}}v k$ to an intermediary flat band. The two gaps at $k=0$ are given by $\Delta$. (a) Plot of the response function for $\Delta=3$, $v=g=1$, and $\eta=0.01$, with the band structure plotted in inset. For resonant driving between the band edges of the upper and lower bands (i.e. $\hbar\omega=2\Delta$), the response is entirely captured by the quadrupolar part. (b) Plots of the response function with fixed $\Delta=3$ and $g=1$ for four different choices of $v/g$, which crucially tunes the extrinsic quantum geometry of the upper/lower bands at $k=0$.}
    \label{fig:minimal model}
\end{figure}

\paragraph{Minimal model} A minimal lattice model of tightly-bound orbitals exhibiting the quadrupolar photovoltaic effect must have at least three orbitals in the unit cell. Furthermore, not all three-band models are sufficient for a quadrupolar response: if for every pair of resonantly driven bands the Bloch states are well described by an effective $\mathbb{CP}^1$ Hilbert space, then the response will still vanish. We thus seek a model in which three bands admix in a way such that the Bloch states explore a nontrivial subspace of $\mathbb{CP}^{n>1}$.

For the purpose of analytical simplicity, we first study the set of three-orbital Bloch Hamiltonians
\begin{align}
    h(\mathbf{k}) = \hspace{-0.1cm}\left[ \hspace{-0.1cm} \begin{array}{ccc} d_3(\mathbf{k}) & \frac{d_1(\mathbf{k}) - id_2(\mathbf{k})}{\sqrt{2}} & -id_4(\mathbf{k}) \\
    \frac{d_1(\mathbf{k}) + id_2(\mathbf{k})}{\sqrt{2}} & 0 & \frac{-d_1(\mathbf{k}) - id_2(\mathbf{k})}{\sqrt{2}} \\
    id_4(\mathbf{k}) & \frac{-d_1(\mathbf{k}) + id_2(\mathbf{k})}{\sqrt{2}} & -d_3(\mathbf{k})
    \end{array} \hspace{-0.1cm} \right]
\end{align}
which anticommute with the chiral symmetry operator $S = [[0,0,1],[0,1,0],[1,0,0]]$. This guarantees a flat band at zero energy $\epsilon_{0,\mathbf{k}} = 0$ as well as two conjugate bands $\epsilon_{\pm,\mathbf{k}} = \pm |\mathbf{d}(\mathbf{k})|$. The Bloch states can be analytically written as smooth functions of the unit vector $\hat{d}(\mathbf{k}) = \mathbf{d}(\mathbf{k}) / |\mathbf{d}(\mathbf{k})|$ on $S^3$ and, importantly, explore a subspace of $\mathbb{CP}^{2}$ that is \textit{not} $\mathbb{CP}^1$. We consider the 1D model specified by $d_1(k) = 0$, $d_2(k) = v \sin ka$ , $d_3(k) = \Delta$, $d_4(k) = g (2-2\cos ka)$ where $a=1$ is the lattice constant. This corresponds to a centrosymmetric but $\mathcal{T}$-breaking system that allows one to study a linearly-polarized photogalvanic effect (LPGE) at linear order in $q$ \cite{lu2025}. When both $v$ and $g$ are nonzero, all three bands admix with nontrivial non-Abelian Berry connections as a function of $k$, allowing the states of each pair of bands to have nontrivial variation in the Hilbert space direction associated with the third band.

Fig. \ref{fig:minimal model} illustrates the band structure and linear-in-$q$ shift response for this model, placing the chemical potential between bands 2 and 3. When $\omega=2\Delta/\hbar$, the magnitude of the response is dominantly controlled by the quadrupolar contribution. This is because the remaining contribution, given by Eq. (\ref{sigma ene}), vanishes when resonantly driving two bands that are energetically particle-hole symmetric, i.e. when $v_{mk}+v_{nk}=0$. The response is therefore determined purely by the quadrupolar part of Eq. (\ref{sigma mul}), where one can show analytically (dropping spatial indices)
\begin{equation}
    \left.\mathcal{I}_{13,k,(Q)}\right|_{k=0} = -\frac{gv^2}{2\Delta^3}.
\end{equation}
This quantity is the sole contribution to the $k$-integral determining the $\omega=2\Delta/\hbar$ quadrupolar conductivity in the clean limit when $\Delta$ is larger than the bandwidth of the dispersing bands. The fact that it vanishes when either $v$ or $g$ is zero reflects its reliance on the mutual admixture of all three bands as a function of $k$, i.e. on the resonantly driven bands having quantum geometry extrinsic to their two-state subspace.

\begin{figure}
    \centering
    \hspace{0.25in}\includegraphics[width=\linewidth]{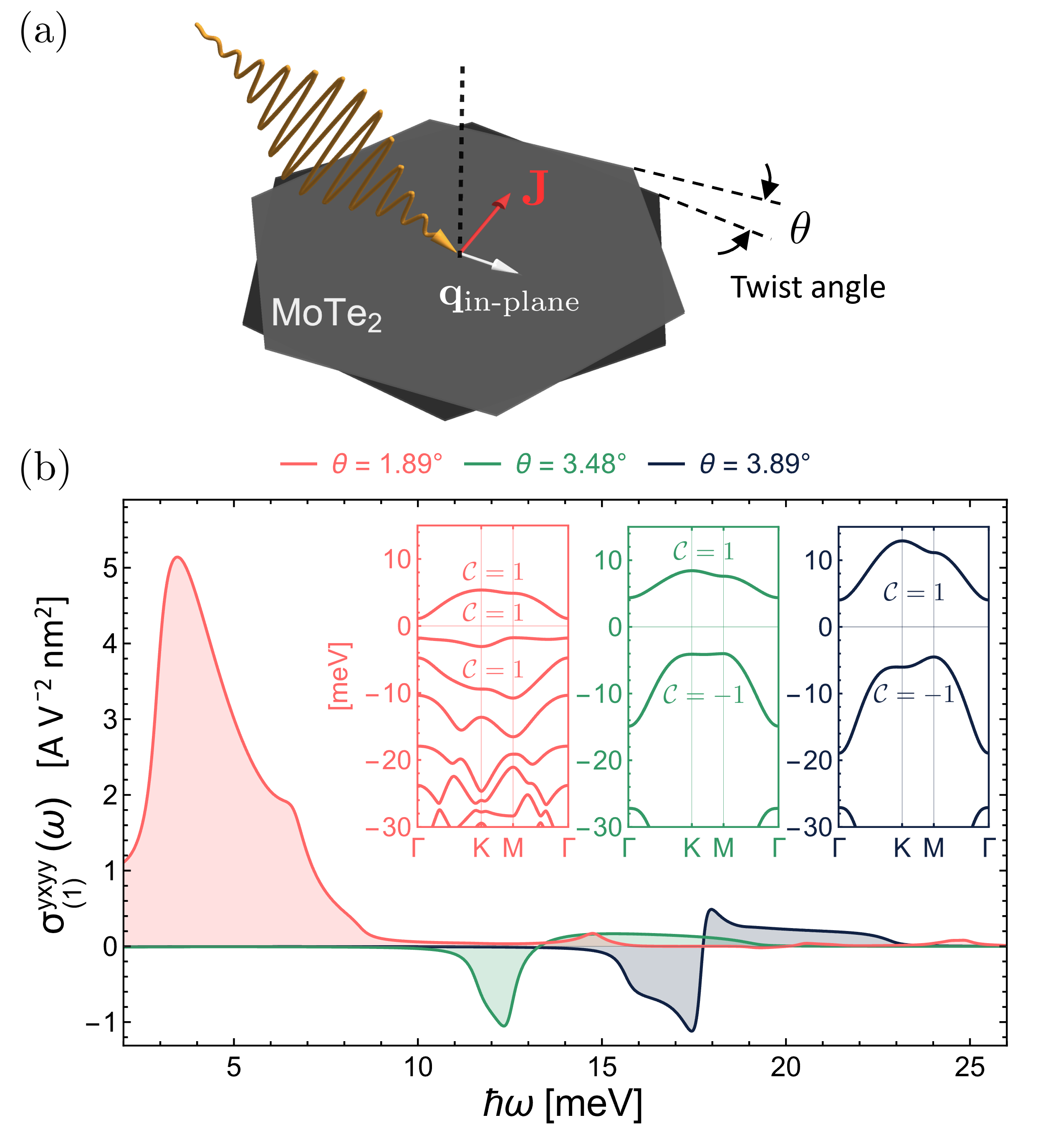}
    \caption{Oblique-incidence photocurrents probe the extrinsic quantum geometry of the moir\'e bands in tMoTe$_2$. (a) Schematic of a photocurrent experiment in a twisted bilayer material whereby oblique incidence introduces an in-plane component to the optical wavevector. (b) The $yxyy$ component of the real linear-in-wavevector photovoltaic conductivity (for which the magnetic dipole contribution is forbidden by symmetry) computed from continuum models of valley-and-spin-polarized tMoTe$_2$ at three different twist angles. The considerably larger response for the twist angle where the top few bands have equal Chern number can be anticipated from the viewpoint of extrinsic quantum geometry.}
    \label{fig:tMoTe2}
\end{figure}

\paragraph{Oblique photocurrents in twisted TMD bilayers} The  principles outlined in this paper have striking implications for the photocurrent response of twisted bilayers, a recent  area of active interest \cite{kaplan2022,lu2025,chaudhary2022,chen2024enhancing,kumar2024terahertzphotocurrentprobequantum}. In particular, twisted bilayer molybdenum ditelluride (tMoTe$_2$) has attracted much attention due to evidence that it exhibits multiple consecutive flat moir\'e Chern bands of the \textit{same} Chern number $\mathcal{C}=1$ at small twist angles \cite{Wu2018,xu2023,zhang2025}, which opens exciting possibilities for non-Abelian fractional Chern insulating states \cite{chen2025,xu2025,li2026}. Since two consecutive bands of equal Chern number \textit{necessarily} cannot be represented by regularized two-band model, this implies that more than two bands are strongly admixing in tMoTe$_2$, making it an excellent target for probing extrinsic quantum geometry through the quadrupolar photocurrent response.

In Fig. \ref{fig:tMoTe2} we report the terahertz quadrupolar photocurrent response calculated for representative twist angles in twisted MoTe\textsubscript{2}. In these systems, an emergent inversion symmetry importantly guarantees the absence of a conventional BPVE, 
suggesting that a terahertz photocurrent experiment on tMoTe$_2$ can serve as a direct probe of the extrinsic quantum geometry of its band structure. At larger angles $3.89^\circ$ and $3.48^\circ$, the topmost moir\'e valence bands carry opposite (spin) Chern numbers $\mathcal{C}=+1,-1$, which we model using a DFT-derived continuum model \cite{Wang2024}. We consider filling $\nu = 1$  with one hole per moir\'e unit cell; here, twisted MoTe$_2$ is spin-polarized with a filled topmost moir\'e band \cite{Wang2024}, permitting a quadrupolar BPVE $\sigma_{\textrm{(1),PV}}^{yxyy}$ for linearly-polarized light, and serving as a sensitive probe to spontaneous time-reversal symmetry breaking.

In contrast, the quadrupolar BPVE shows a striking increase at small twist angles $1.89^{\circ}$. Here, prior theoretical and experimental works indicate a sequence of valence bands with \textit{same} Chern numbers $\mathcal{C} = +1$ due to lattice relaxation effects, which we model via the DFT-derived continuum model of Ref. \cite{Cheng2025}. Consequently, an interband transition between the top two valence bands \textit{cannot} be described within a two-band space alone, leading to a drastically enhanced electric quadrupole and anomalous velocity coupling from extrinsic quantum geometry. We note that the magnetic dipole contribution vanishes by symmetry for $\sigma_{(1),\textrm{PV}}^{yxyy}$. Resonantly driving the top two valence bands via an oblique-incidence THz field therefore directly probes extrinsic quantum geometry across a topological transition. Strikingly, the magnitude of $|\mathbf{q}|\sigma^{\nu\mu\alpha\beta}_{(1),\text{PV}}$ is comparable to the magnitude of electric-dipole shift conductivities predicted for inversion-breaking 2D platforms (see Supplementary Material), further suggesting that beyond-dipole effects are generally important in moir\'e systems \cite{lu2025}.

In summary, our reformulation of the wavevector-dependent BPVE (the photon drag effect) at leading order in the wavevector provides a conceptual bridge between multipolar corrections and band-geometric frameworks for nonlinear optical responses, and it highlights a previously neglected contribution representing a purely quantum geometric mechanism for the breakdown of the dipole approximation. Further clarifying the role of band geometry in spatially dispersive optics, particularly their implications for nonlinear optical response functions in moir\'e materials, is an exciting direction for future work.
\\
\\
\noindent \emph{Acknowledgements} We thank Christophe De Beule, Gene Mele, Stefan Divic, Jixun K. Ding, and Deven Carmichael for helpful discussions. S.G. and M.C. acknowledge support from the U.S.\ Department of Energy, Office of Basic Energy Sciences, under Awards No.\ DE-FG02-84ER45118 (S.G.) and DE-SC0024494 (M.C.). S.G. additionally acknowledges support from the NSF GRFP. S.C acknowledges support from JSPS KAKENHI (No. JP23H04865), MEXT, Japan.
 This research was supported in part by grant NSF PHY-2309135 and the Gordon and Betty Moore Foundation Grant No. 2919.02 to the Kavli Institute for Theoretical Physics (KITP).

\bibliography{references}

\appendix

\section*{Appendix}

\paragraph{Appendix A: Note on finite-wavevector current operators} One must take care in general when defining the finite-wavevector current operator $j^\mu_\mathbf{q}$, as pointed out in \cite{mckay2024}. Often-used current operators in effective models can be easily shown to violate the continuity equation. In the case of the photon-drag conductivity, one can derive the following part of the response directly from charge conservation,
\begin{equation}
\begin{split}
    &\frac{q_\alpha q_\beta}{\omega^2} \mathcal{M}^{\mu\alpha\beta}_{mk',nk} = \\
    &~~~~~~~\langle u_{n\mathbf{k}}|u_{m\mathbf{k}'}\rangle\left(i\nabla^\mu_\mathbf{k}+\mathcal{A}^\mu_{m\mathbf{k}'}-\mathcal{A}^\mu_{n\mathbf{k}}\right)\langle u_{m\mathbf{k}'} | u_{n\mathbf{k}}\rangle \\
    &~~~~~~~+\tau|\langle u_{m\mathbf{k}} | u_{n\mathbf{k}} \rangle|^2(v^\mu_{m\mathbf{k}'}-v^\mu_{n\mathbf{k}}).
\end{split}
\end{equation}
Note that this does \textit{not} agree in general with expressions for the photon-drag conductivity in the literature, which implicitly assume $\langle u_{m,\mathbf{k}+\frac{\mathbf{q}}{2}}|u_{n,\mathbf{k}-\frac{\mathbf{q}}{2}}\rangle = i\mathbf{q}\cdot\langle u_{m,\mathbf{k}+\frac{\mathbf{q}}{2}}|\mathbf{r}|u_{n,\mathbf{k}-\frac{\mathbf{q}}{2}}\rangle$. Deviations from this assumption are of fundamental interest to the study of the photon-drag effect and are an important direction for future work.

\end{document}


\title{Supplementary material: \\ Extrinsic quantum geometry in the quadrupolar bulk photovoltaic effect}
\author{Steven Gassner, Swati Chaudhary, Martin Claassen}
\date{\today}

\maketitle

\tableofcontents

\section{Diagrammatic perturbation theory}

In this section, we derive the second-order optical conductivity in the velocity gauge augmented with wavevector dependence, following the scheme of Parker et al \cite{parker2019}. For compactness, we use a ``space-time" notation,
\begin{equation}
\begin{split}
    q_j & \equiv (i\Omega^{(j)}_m,\mathbf{q}_j) ~~~~\text{(photons)} \\
    k & \equiv (i\omega_n,\mathbf{k}) ~~~~\text{(electrons)}
\end{split}
\end{equation}
to keep track of both the (Matsubara) frequency and momenta of the photons and the electrons. The index $j$ runs through the total number of photons, $N+1$ in the case of an $N^\text{th}$-order response. Because we restrict to noninteracting electrons, the formulas will be generated by summing over a single unconstrained fermionic Matsubara frequency $i\omega_n$ from a single fermion loop. We will then analytically continue the bosonic Matsubara frequencies $i\Omega^{(j)}_m \to \omega_j + i\eta$ to obtain the response function.

We define the wavevector-dependent second-order optical conductivity as follows,
\begin{equation}
    \left\langle j^\mu(Q)\right\rangle = \sigma^{\mu\alpha\beta}\left(Q;q_1,q_2\right) E_\alpha(q_1)E_\beta(q_2)
\end{equation}
where $Q = q_1 + q_2$. We choose the following electromagnetic gauge choice when writing the Hamiltonian,
\begin{equation}
    \mathcal{H}_\mathbf{A} = \frac{\left[-i\hbar\boldsymbol{\nabla} + e\mathbf{A}(\mathbf{r},t)\right]^2}{2m_\text{e}} + V(\mathbf{r})
\end{equation}
where $V(\mathbf{r})$ is the static periodic potential. That is, we set the scalar potential $\phi(\mathbf{r},t)$ to zero. We complement this analysis in Section \ref{scalar potential gauge} in which we consider the response in one dimension to an electric field arising purely from a spatially varying scalar potential.

Formally expanding the Hamiltonian in powers of $\mathbf{A}(\mathbf{r},t) = \sum_{\mathbf{q}_j} \mathbf{A}(\omega_j,\mathbf{q}_j) e^{i(\mathbf{q}_j\cdot\mathbf{r}-\omega_j t)} + \text{c.c.}$ yields the wavevector-dependent current operators,
\begin{equation}
    \mathcal{H}_\mathbf{A} = \mathcal{H}_0 + e\sum_{\mathbf{q}_j} j^\alpha_{\mathbf{q}_j} A_\alpha(\omega_j,\mathbf{q}_j) + e^2\sum_{\mathbf{q}_j,\mathbf{q}_j'}j^{\alpha\beta}_{\mathbf{q}_j,\mathbf{q}_j'}A_\alpha(\omega_j,\mathbf{q}_j)A_\beta(\omega'_j,\mathbf{q}_j') + \hdots
\end{equation}
The $N^\text{th}$-order optical conductivity involves all current operators up to $N+1$ order in the gauge field.

The second-order optical conductivity takes the form,
\begin{equation}
    \sigma^{\mu\alpha\beta}(Q;q_1,q_2) = \frac{1}{-i\Omega^{(1)}_mi\Omega^{(2)}_m}\frac{-e^3}{\hbar^2}\left(\text{[Diagram 1]+[Diagram 2]+[Diagram 3]+[Diagram 4]}\right)
\end{equation}
summarized in the following four diagrams,
\begin{center}
\includegraphics[width=0.6\linewidth]{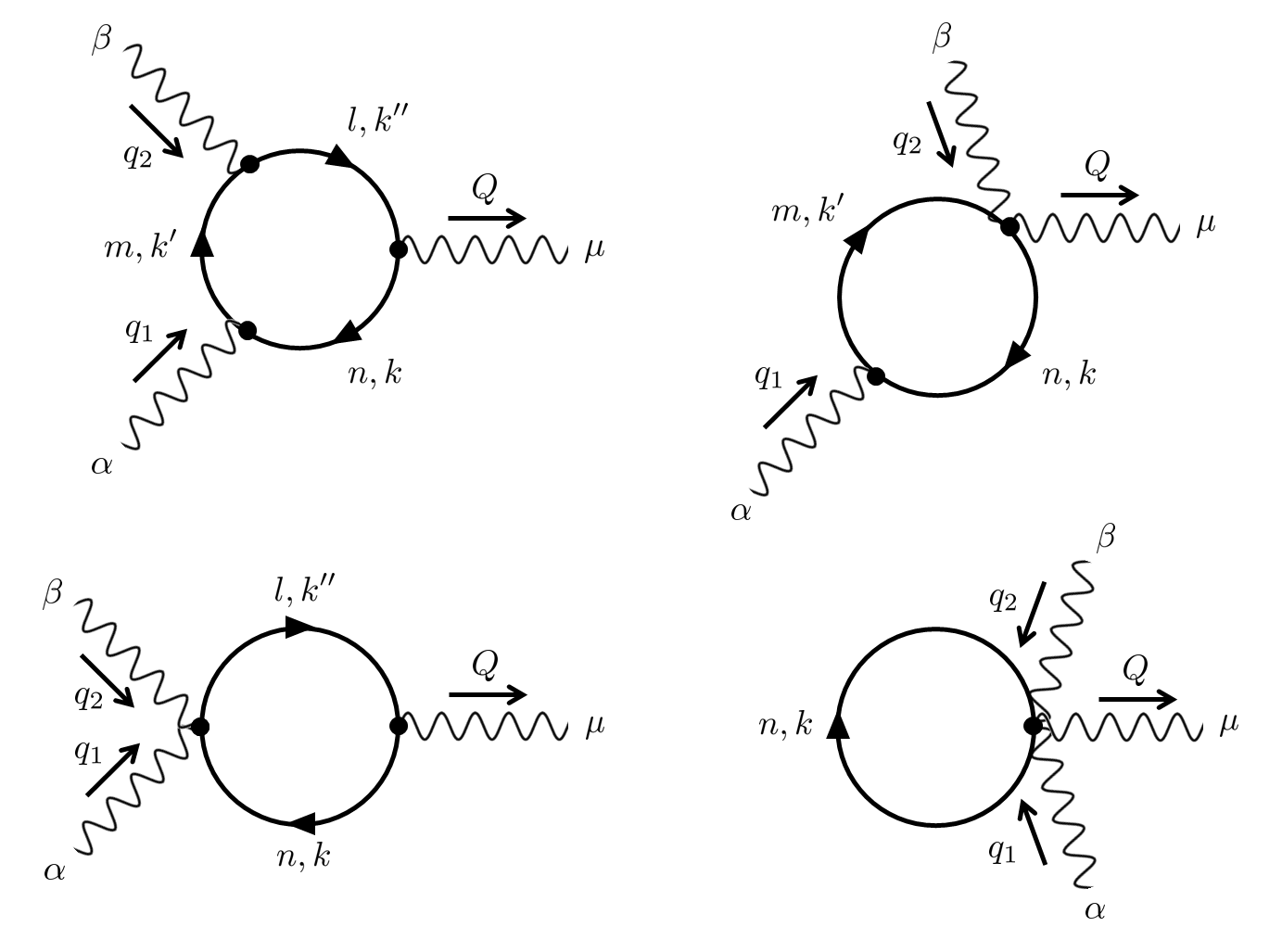}
.
\end{center}
Each diagram corresponds schematically to one of the following expressions,
\begin{equation}
\begin{split}
    \text{tr}\left\{    
    G_{k''} j^\beta_{q_2} G_{k'} j^\alpha_{q_1} G_{k} j^\mu_{-Q}
    \right\}
    ~~~~~&~~~~~
    \text{tr}\left\{G_{k'} j^\alpha_{q_1} G_k j^{\mu\beta}_{-Q,q_2}\right\} \\
    \text{tr}\left\{G_{k''} \frac{1}{2}j^{\alpha\beta}_{q_1,q_2} G_{k} j^{\mu}_{-Q}\right\}~~~~~~&~~~~~~\text{tr}\left\{G_k\frac{1}{2}j^{\mu\alpha\beta}_{-Q,q_1,q_2}\right\}
\end{split}
\end{equation}
Here, $G_k = \sum_n\frac{|\psi_{n\mathbf{k}}\rangle\langle \psi_{n\mathbf{k}}|}{i\omega_n-\omega_{n\mathbf{k}}}$ is the fermionic Green's function expressed in terms of the eigenstates and eigenvalues of the unperturbed Bloch Hamiltonian $h_\mathbf{k}|\psi_{n\mathbf{k}}\rangle = \hbar\omega_{n\mathbf{k}}|\psi_{n\mathbf{k}}\rangle$, where $h_\mathbf{k} = e^{-i\mathbf{k}\cdot\mathbf{r}}\mathcal{H}_0 e^{i\mathbf{k}\cdot\mathbf{r}}$ \footnote{We trust the reader not to confuse the ``$n$'' indexing the imaginary Matsubara frequency $i\omega_n$ and the band index ``$n$'' in the band energy $\hbar\omega_{n\mathbf{k}}$.}. The trace schematically denotes summing over internal fermion flavors (band indices) and the single unconstrained fermion momentum and fermion Matsubara frequency. Momentum and energy conservation is enforced at each vertex. The factors of $1/2$ account for the fact that the response function should be symmetrized over $[\alpha,q_1]\leftrightarrow[\beta,q_2]$, so the diagrams that already have this symmetry should not be double-counted.

For our purposes (the photovoltaic response $q_1 = -q_2$), only the top two diagrams will be relevant for the resonant behavior of the response. The bottom two diagrams are only important for regularizing the limit $q_1,q_2\to 0$. We will evaluate the top two diagrams in turn. From a pedagogical standpoint, the diagrams are in reverse order, so we will consider Diagram 2 before Diagram 1.

\subsection*{Diagram 2}
\begin{center}
    \includegraphics[height=2in]{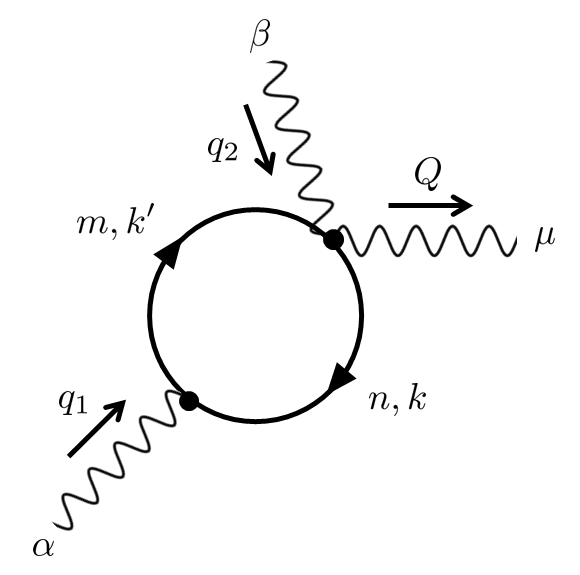}
\end{center}
Starting at the ``$\mu$" vertex and following the loop clockwise, we can write down the trace corresponding to this diagram as,
\begin{align}
    [\text{Diagram 2}] &= \frac{1}{\beta L^{2d}}\sum_{i\omega_n} \sum_{\mathbf{k},\mathbf{k}'} \delta_{\mathbf{k}'-\mathbf{k}-\mathbf{q}_1} \\
    &~~~~~~~~~~~\times\sum_{nm} \text{tr}\left\{ \frac{|\psi_{m\mathbf{k}'}\rangle\langle\psi_{m\mathbf{k}'}|}{i\omega_n+i\Omega^{(1)}_m-\omega_{m\mathbf{k}'}} j^\alpha_{q_1} \frac{|\psi_{n\mathbf{k}}\rangle\langle\psi_{n\mathbf{k}}|}{i\omega_n-\omega_{n\mathbf{k}}} j^{\mu\beta}_{-Q,q_2}\right\} \\
    &= \frac{1}{\beta L^{2d}}\sum_{i\omega_n} \sum_{\mathbf{k},\mathbf{k}'}^\sim\sum_{nm}\frac{j^\alpha_{m\mathbf{k}',n\mathbf{k}}j^{\mu\beta}_{n\mathbf{k},m\mathbf{k}'}}{(i\omega_n-\omega_{n\mathbf{k}})(i\omega_n+i\Omega^{(1)}_m - \omega_{m\mathbf{k}'})}
\end{align}
where $j^\alpha_{m\mathbf{k}',n\mathbf{k}} = \langle \psi_{m\mathbf{k}'} | j^\alpha_{\mathbf{q}_1} | \psi_{n\mathbf{k}}\rangle$, $~j^{\mu\beta}_{n\mathbf{k},m\mathbf{k}'} = \langle \psi_{n\mathbf{k}} | j^{\mu\beta}_{-\mathbf{Q},\mathbf{q}_2} | \psi_{m\mathbf{k}'}\rangle$ and we absorb the momentum-conserving delta function in the restricted sum $\tilde{\sum}_{\mathbf{k},\mathbf{k}'}$. One can choose to sum over one of these momenta to write everything in terms of one unconstrained fermion momentum boosted by combinations of $\mathbf{q}_1, \mathbf{q}_2$. However, we delay this step in order to delay the choice of parameterizing the arguments of the Bloch states and energies in terms of the photon momenta $\mathbf{q}_j$. Flexibility in this regard will be helpful when expanding response functions in powers of $\mathbf{q}$.

We now perform the sum over the unconstrained fermionic Matsubara frequency $i\omega_n$. The general procedure is to perform a partial-fraction decomposition of the summand and then to use the fact that $\frac{1}{\beta}\sum_{i\omega_n}\frac{1}{i\omega_n-\omega_{n\mathbf{k}}} = f_{n\mathbf{k}}$ where $f_{n\mathbf{k}}\equiv f_\text{FD}(\hbar\omega_{n\mathbf{k}})$ is the Fermi-Dirac distribution evaluated at $\hbar\omega_{n\mathbf{k}}$. Importantly, this equality also remains true when $i\omega_n$ is shifted by any bosonic Matsubara frequency $i\Omega_m$.

The useful partial-fraction decomposition identity for Diagram 2 is,
\begin{equation}
    \frac{1}{(x-A)(x-B)} = \frac{[A-B]^{-1}}{x-A} + \frac{[B-A]^{-1}}{x-B} = \frac{1}{A-B}\left(\frac{1}{x-A}-\frac{1}{x-B}\right)
\end{equation}
Partial fraction decomposition yields,
\begin{align}
    [\text{Diagram 2}] = \frac{1}{\beta L^{2d}}\sum_{i\omega_n}\sum_{nm}\sum_{\mathbf{k},\mathbf{k}'}^\sim j^{\mu\beta}_{n\mathbf{k},m\mathbf{k}'}j^\alpha_{m\mathbf{k}',n\mathbf{k}}\frac{1}{\omega_{n\mathbf{k}}+i\Omega^{(1)}_m-\omega_{m\mathbf{k}}}\left(\frac{1}{i\omega_n-\omega_{n\mathbf{k}}}-\frac{1}{i\omega_n+i\Omega^{(1)}_m-\omega_{m\mathbf{k}'}}\right)
\end{align}
which yields, after performing the Matsubara sum and analytically continuing $i\Omega^{(1)}_m \to \omega_1 + i\eta$
\begin{equation}
    [\text{Diagram 2}] = \frac{1}{L^{2d}}\sum_{nm}\sum_{\mathbf{k},\mathbf{k}'}^\sim j^{\mu\beta}_{n\mathbf{k},m\mathbf{k}'}j^\alpha_{m\mathbf{k}',n\mathbf{k}}\frac{f_{n\mathbf{k}}-f_{m\mathbf{k}'}}{\omega_{n\mathbf{k}}-\omega_{m\mathbf{k}'}+\omega_1+i\eta}
    .
\end{equation}
We will often abbreviate $f_{n\mathbf{k},m\mathbf{k}'} \equiv f_{n\mathbf{k}}-f_{m\mathbf{k}'}$ and $\omega_{n\mathbf{k},m\mathbf{k}'}=\omega_{n\mathbf{k}}-\omega_{m\mathbf{k}'}$.

For our purposes, we are interested in the photovoltaic response given by $q_1=-q_2=q$. Restoring the abbreviated sums over momenta and remembering to symmetrize over $[(\alpha,q_1)\leftrightarrow (\beta,q_2)]$, we obtain
\begin{equation}
    [\text{Diagram 2}] = \frac{1}{L^{2d}}\sum_{nm}\sum_{\mathbf{k},\mathbf{k}'}\delta_{\mathbf{k}'-\mathbf{k}-\mathbf{q}} \left(j^{\mu\beta}_{n\mathbf{k},m\mathbf{k}'}j^\alpha_{m\mathbf{k}',n\mathbf{k}}\frac{f_{n\mathbf{k},m\mathbf{k}'}}{\omega_{n\mathbf{k},m\mathbf{k}'}+\omega+i\eta} + j^{\mu\alpha}_{m\mathbf{k}',n\mathbf{k}}j^\beta_{n\mathbf{k},m\mathbf{k}'}\frac{f_{m\mathbf{k}',n\mathbf{k}}}{\omega_{m\mathbf{k}',n\mathbf{k}}-\omega+i\eta}\right)
\end{equation}
where in the second term we implement the substitution $\mathbf{q}\to -\mathbf{q}$ by swapping the summed momenta $\mathbf{k}\leftrightarrow\mathbf{k}'$, and we additionally swap the dummy indices $m\leftrightarrow n$ to keep $n$ associated with $\mathbf{k}$ and $m$ associated with $\mathbf{k}'$.

Observe finally that the second term is precisely the complex conjugate of the first, up to an exchange of $\alpha\leftrightarrow\beta$. This means the contribution to the response from Diagram 2 takes the following form,
\begin{equation}
    [\text{Diagram 2}] = \frac{1}{L^{2d}}\sum_{\mathbf{k},\mathbf{k}'}\delta_{\mathbf{k}'-\mathbf{k}-\mathbf{q}}\sum_{nm}f_{m\mathbf{k}',n\mathbf{k}}\left(\frac{-i\left[\mathcal{M}^{\mu\alpha\beta}_{m\mathbf{k}',n\mathbf{k}}\right]_{(2)}}{\omega_{m\mathbf{k}',n\mathbf{k}}-\omega+i\eta}+\frac{i\left[\mathcal{M}^{\mu\beta\alpha}_{m\mathbf{k}',n\mathbf{k}}\right]^*_{(2)}}{\omega_{m\mathbf{k}',n\mathbf{k}}-\omega-i\eta}\right)
\end{equation}
\begin{equation}
    \left[\mathcal{M}^{\mu\alpha\beta}_{m\mathbf{k}',n\mathbf{k}}\right]_{(2)} = ij^\beta_{n\mathbf{k},m\mathbf{k}'}j^{\mu\alpha}_{m\mathbf{k}',n\mathbf{k}}
    .
\end{equation}

\subsection*{Diagram 1}
\begin{center}
    \includegraphics[width=0.3\linewidth]{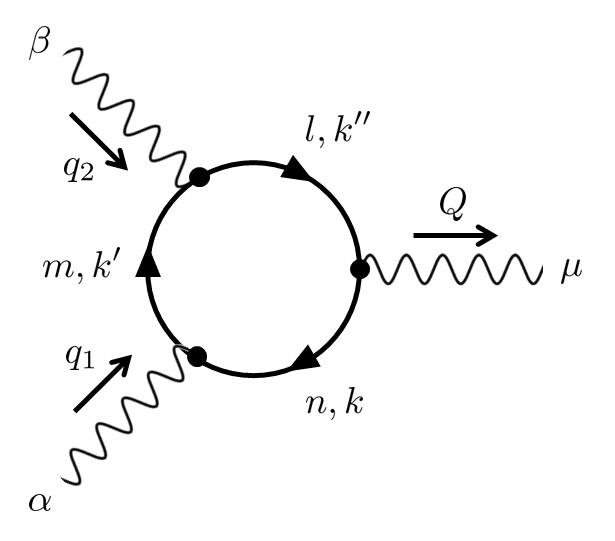}
\end{center}

Starting at the ``$\mu$" vertex and following the loop clockwise, we can write down the trace corresponding to this diagram as,
\begin{align}
    [\text{Diagram 1}] &= \frac{1}{\beta L^{3d}}\sum_{i\omega_n} \sum_{\mathbf{k},\mathbf{k}',\mathbf{k}''} \delta_{\mathbf{k}'-\mathbf{k}-\mathbf{q}_1}\delta_{\mathbf{k}''-\mathbf{k}'-\mathbf{q}_2} \\
    &~~~~~~~~~~~\times\sum_{nml} \text{tr}\left\{\frac{|\psi_{l\mathbf{k}''}\rangle\langle\psi_{l\mathbf{k}''}|}{i\omega_n+i\Omega^{(1,2)}_m-\omega_{l\mathbf{k}''}} j^\beta_{q_2} \frac{|\psi_{m\mathbf{k}'}\rangle\langle\psi_{m\mathbf{k}'}|}{i\omega_n+i\Omega^{(1)}_m-\omega_{m\mathbf{k}'}} j^\alpha_{q_1} \frac{|\psi_{n\mathbf{k}}\rangle\langle\psi_{n\mathbf{k}}|}{i\omega_n-\omega_{n\mathbf{k}}} j^\mu_{-Q}\right\} \\
    &= \frac{1}{\beta L^{3d}}\sum_{i\omega_n} \sum_{\mathbf{k},\mathbf{k}',\mathbf{k}''}^\sim\sum_{nml}\frac{j^\beta_{l\mathbf{k}'',m\mathbf{k}'}j^\alpha_{m\mathbf{k}',n\mathbf{k}}j^\mu_{n\mathbf{k},l\mathbf{k}''}}{(i\omega_n-\omega_{n\mathbf{k}})(i\omega_n+i\Omega^{(1)}_m - \omega_{m\mathbf{k}'})(i\omega_n+i\Omega^{(1,2)}_m-\omega_{l\mathbf{k}''})}
\end{align}
where $i\Omega^{(1,2)}_m = i\Omega^{(1)}_m+i\Omega^{(2)}_m$, $~j^\alpha_{m\mathbf{k}',n\mathbf{k}} = \langle \psi_{m\mathbf{k}'} | j^\alpha_{\mathbf{q}_1} | \psi_{n\mathbf{k}}\rangle$, and we absorb the momentum-conserving delta functions in the restricted sum $\tilde{\sum}_{\mathbf{k},\mathbf{k}',\mathbf{k}''}$. One can choose to sum over two of these momenta to write everything in terms of one unconstrained fermion momentum boosted by combinations of $\mathbf{q}_1, \mathbf{q}_2$. However, we delay this step in order to delay the choice of parameterizing the arguments of the Bloch states and energies in terms of the photon momenta $\mathbf{q}_j$. Flexibility in this regard will be helpful when expanding response functions in powers of $\mathbf{q}$.

We now perform the sum over the unconstrained fermionic Matsubara frequency $i\omega_n$. The general procedure is to perform a partial-fraction decomposition of the summand and then to use the fact that $\frac{1}{\beta}\sum_{i\omega_n}\frac{1}{i\omega_n-\omega_{n\mathbf{k}}} = f_{n\mathbf{k}}$ where $f_{n\mathbf{k}}\equiv f_\text{FD}(\hbar\omega_{n\mathbf{k}})$ is the Fermi-Dirac distribution evaluated at $\hbar\omega_{n\mathbf{k}}$. Importantly, this equality also remains true when $i\omega_n$ is shifted by any bosonic Matsubara frequency $i\Omega_m$.

The useful partial-fraction decomposition identity for Diagram 1 is,
\begin{equation}
    \frac{1}{(x-A)(x-B)(x-C)} = \frac{[(A-B)(A-C)]^{-1}}{x-A} + \frac{[(B-A)(B-C)]^{-1}}{x-B} + \frac{[(C-A)(C-B)]^{-1}}{x-C}
    .
\end{equation}
Let us abbreviate $a\equiv (n,\mathbf{k})$, $b\equiv(m,\mathbf{k}')$, and $c\equiv(l,\mathbf{k}'')$, and write $\epsilon_a = \omega_{n\mathbf{k}}$, $\epsilon_b = \omega_{m\mathbf{k}'}$, and $\epsilon_c = \omega_{l\mathbf{k}''}$. Partial fraction decomposition yields,
\begin{equation}
\begin{split}
    [\text{Diagram 1}] = \frac{1}{\beta}\sum_{i\omega_n}\sum_{a,b,c}j^\mu_{ca}j^\alpha_{ab}j^\beta_{bc}&\left(\frac{[(\epsilon_a+i\Omega^{(1)}_m-\epsilon_b)(\epsilon_a+i\Omega^{(1,2)}_m-\epsilon_c)]^{-1}}{i\omega_n-\epsilon_a}\right. \\
    &+ \frac{[(\epsilon_b-i\Omega^{(1)}_m-\epsilon_a)(\epsilon_b-i\Omega^{(1)}_m+i\Omega^{(1,2)}_m-\epsilon_c)]^{-1}}{i\omega_n+i\Omega^1_m-\epsilon_b} \\
    &+\left.\frac{[(\epsilon_c-i\Omega^{(1,2)}_m-\epsilon_a)(\epsilon_c-i\Omega^{(1,2)}_m+i\Omega_m^{(1)}-\epsilon_b)]^{-1}}{i\omega_n+i\Omega-\epsilon_c}\right)
\end{split}
\end{equation}
which yields, after performing the Matsubara sum and analytically continuing $i\Omega^{(j)}_m \to \omega_j + i\eta$, $i\Omega^{(1)}_m+i\Omega^{(2)}_m \to \Omega+i\eta$ \footnote{Introducing a finite relaxation rate $\eta$ must be done in such a way that all input frequencies (regardless of sign!) receive the same imaginary shift $\omega_j + i\eta$. See \cite{parker2019} for more details.}
\begin{equation}
\begin{split}
    [\text{Diagram 1}] = \sum_{a,b,c}j^\mu_{ca}j^\alpha_{ab}j^\beta_{bc}&\bigg(\frac{f_a}{(\epsilon_{ab}+\omega_1 + i\eta)(\epsilon_{ac}+\Omega+2i\eta)}\\
    &+ \frac{f_b}{(\epsilon_{ba}-\omega_1-i\eta)(\epsilon_{bc}+\omega_2+i\eta)}\\
    &+\frac{f_c}{(\epsilon_{ca}-\Omega-2i\eta)(\epsilon_{cb}-\omega_2-i\eta)}\bigg).
\end{split}
\end{equation}
As written, the summand does not manifestly vanish when $f_a=f_b=f_c$. One may worry about divergences / numerical instability when summing over cases like this. It is advantageous to express it as follows (where $f_{ab}\equiv f_a-f_b$)
\begin{equation}
    [\text{Diagram 1}] = \sum_{a,b,c}j^\mu_{ca}j^\alpha_{ab}j^\beta_{bc}\frac{f_{ab}(\epsilon_{bc}+\omega_2+i\eta)+f_{cb}(\epsilon_{ab}+\omega_1+i\eta)}{(\epsilon_{ab}+\omega_1+i\eta)(\epsilon_{bc}+\omega_2+i\eta)(\epsilon_{ac}+\Omega+2i\eta)}
    .
\end{equation}
Finally, for our purposes, we consider the case $q_1 = -q_2 = q$. Restoring abbreviations and remembering to symmetrize over $[(\alpha,q)\leftrightarrow (\beta,-q)]$, we obtain
\begin{align}
    [\text{Diagram 1}] &= \sum_{\mathbf{k},\mathbf{k}'}\delta_{\mathbf{k}'-\mathbf{k}-\mathbf{q}}\sum_{nml}\frac{j^\beta_{l\mathbf{k},m\mathbf{k}'}j^\alpha_{m\mathbf{k}',n\mathbf{k}}j^\mu_{n\mathbf{k},l\mathbf{k}}}{\omega_{n\mathbf{k},l\mathbf{k}}+2i\eta}\left(\frac{f_{n\mathbf{k},m\mathbf{k}'}}{\omega_{n\mathbf{k},m\mathbf{k}'}+\omega+i\eta}+\frac{f_{l\mathbf{k},m\mathbf{k}'}}{\omega_{m\mathbf{k}',l\mathbf{k}}-\omega+i\eta}\right) + [(\alpha,q)\leftrightarrow(\beta,-q)] \\
    \begin{split}
        &= \sum_{\mathbf{k},\mathbf{k}'}\delta_{\mathbf{k}'-\mathbf{k}-\mathbf{q}}\sum_{nml}\left\{\frac{j^\beta_{l\mathbf{k},m\mathbf{k}'}j^\alpha_{m\mathbf{k}',n\mathbf{k}}j^\mu_{n\mathbf{k},l\mathbf{k}}}{\omega_{n\mathbf{k},l\mathbf{k}}+2i\eta}\left(\frac{f_{n\mathbf{k},m\mathbf{k}'}}{\omega_{n\mathbf{k},m\mathbf{k}'}+\omega+i\eta}+\frac{f_{l\mathbf{k},m\mathbf{k}'}}{\omega_{m\mathbf{k}',l\mathbf{k}}-\omega+i\eta}\right)\right.\\
        &~~~~~~~~~~~~~~~~~~~~~~~~~~+\left.\frac{j^\alpha_{l\mathbf{k}',n\mathbf{k}}j^\beta_{n\mathbf{k},m\mathbf{k}'}j^\mu_{m\mathbf{k}',l\mathbf{k}'}}{\omega_{m\mathbf{k}',l\mathbf{k}'}+2i\eta}\left(\frac{f_{m\mathbf{k}',n\mathbf{k}}}{\omega_{m\mathbf{k}',n\mathbf{k}}-\omega+i\eta}+\frac{f_{l\mathbf{k}',n\mathbf{k}}}{\omega_{n\mathbf{k},l\mathbf{k}'}+\omega+i\eta}\right)\right\}
        .
    \end{split}
\end{align}
In the second step, we take $\mathbf{q}$ to $-\mathbf{q}$ by swapping the momenta $\mathbf{k}\leftrightarrow\mathbf{k}'$. We additionally swap the dummy indices $m \leftrightarrow n$, to keep $n$ associated with $\mathbf{k}$ and $m$ associated with $\mathbf{k}'$ (band $l$, however, is not relabeled and is now associated with $\mathbf{k}'$ in the bottom set of terms). We can now combine these four terms by relabeling $l\leftrightarrow n$ in the second term and $l\leftrightarrow m$ in the fourth term,
\begin{align}
    \begin{split}
        [\text{Diagram 1}] &= \sum_{\mathbf{k},\mathbf{k}'}\delta_{\mathbf{k}'-\mathbf{k}-\mathbf{q}}\sum_{nml}\left\{\frac{f_{n\mathbf{k},m\mathbf{k}'}}{\omega_{n\mathbf{k},m\mathbf{k}'}+\omega+i\eta}\frac{j^\beta_{l\mathbf{k},m\mathbf{k}'}j^\alpha_{m\mathbf{k}',n\mathbf{k}}j^\mu_{n\mathbf{k},l\mathbf{k}}}{\omega_{n\mathbf{k},l\mathbf{k}}+2i\eta}\right. \\
        &~~~~~~~~~~~~~~~~~~~~~~~~~~+\frac{f_{n\mathbf{k},m\mathbf{k}'}}{\omega_{m\mathbf{k}',n\mathbf{k}}-\omega+i\eta}\frac{j^\beta_{n\mathbf{k},m\mathbf{k}'}j^\alpha_{m\mathbf{k}',l\mathbf{k}}j^\mu_{l\mathbf{k},n\mathbf{k}}}{\omega_{l\mathbf{k},n\mathbf{k}}+2i\eta}\\
        &~~~~~~~~~~~~~~~~~~~~~~~~~~+\frac{f_{m\mathbf{k}',n\mathbf{k}}}{\omega_{m\mathbf{k}',n\mathbf{k}}-\omega+i\eta}\frac{j^\alpha_{l\mathbf{k}',n\mathbf{k}}j^\beta_{n\mathbf{k},m\mathbf{k}'}j^\mu_{m\mathbf{k}',l\mathbf{k}'}}{\omega_{m\mathbf{k}',l\mathbf{k}'}+2i\eta}\\
        &~~~~~~~~~~~~~~~~~~~~~~~~~~\left.+\frac{f_{m\mathbf{k}',n\mathbf{k}}}{\omega_{n\mathbf{k},m\mathbf{k}'}+\omega+i\eta}\frac{j^\alpha_{m\mathbf{k}',n\mathbf{k}}j^\beta_{n\mathbf{k},l\mathbf{k}'}j^\mu_{l\mathbf{k}',m\mathbf{k}'}}{\omega_{l\mathbf{k}',m\mathbf{k}'}+2i\eta}\right\}
    \end{split}
    \\
    \begin{split}
        &= \sum_{\mathbf{k},\mathbf{k}'}\delta_{\mathbf{k}'-\mathbf{k}-\mathbf{q}}\sum_{nml}\left\{\frac{f_{m\mathbf{k}',n\mathbf{k}}}{\omega_{m\mathbf{k}',n\mathbf{k}}-\omega+i\eta}j^\beta_{n\mathbf{k},m\mathbf{k}'}\left(\frac{j^\mu_{m\mathbf{k}',l\mathbf{k}'}j^\alpha_{l\mathbf{k}',n\mathbf{k}}}{\omega_{m\mathbf{k}',l\mathbf{k}'}+2i\eta}-\frac{j^\alpha_{m\mathbf{k}',l\mathbf{k}}j^\mu_{l\mathbf{k},n\mathbf{k}}}{\omega_{l\mathbf{k},n\mathbf{k}}+2i\eta}\right)\right.\\
        &~~~~~~~~~~~~~~~~~~~~~~~~~~+\left.\frac{f_{m\mathbf{k}',n\mathbf{k}}}{\omega_{m\mathbf{k}',n\mathbf{k}}-\omega-i\eta}j^\alpha_{m\mathbf{k}',n\mathbf{k}}\left(\frac{j^\mu_{n\mathbf{k},l\mathbf{k}}j^\beta_{l\mathbf{k},m\mathbf{k}'}}{\omega_{n\mathbf{k},l\mathbf{k}}+2i\eta}-\frac{j^\beta_{n\mathbf{k},l\mathbf{k}'}j^\mu_{l\mathbf{k}',m\mathbf{k}'}}{\omega_{l\mathbf{k}',m\mathbf{k}'}+2i\eta}\right)\right\}
    \end{split}
\end{align}
Observe finally that the second term is precisely the complex conjugate of the first up to an interchange of $\alpha\leftrightarrow\beta$. We can express this as follows,
\begin{equation}
    [\text{Diagram 1}] = \sum_{\mathbf{k},\mathbf{k}'}\delta_{\mathbf{k}'-\mathbf{k}-\mathbf{q}}\sum_{nm}f_{m\mathbf{k},n\mathbf{k}'}\left(\frac{-i\left[\mathcal{M}^{\mu\alpha\beta}_{m\mathbf{k}',n\mathbf{k}}\right]_{(1)}}{\omega_{m\mathbf{k}',n\mathbf{k}}-\omega+i\eta}+\frac{i\left[\mathcal{M}^{\mu\beta\alpha}_{m\mathbf{k}',n\mathbf{k}}\right]^*_{(1)}}{\omega_{m\mathbf{k}',n\mathbf{k}}-\omega-i\eta}\right)
\end{equation}
\begin{equation}
    \left[\mathcal{M}^{\mu\alpha\beta}_{m\mathbf{k}',n\mathbf{k}}\right]_{(1)} = ij^\beta_{n\mathbf{k},m\mathbf{k}'}\sum_l\left(\frac{j^\mu_{m\mathbf{k}',l\mathbf{k}'}j^\alpha_{l\mathbf{k}',n\mathbf{k}}}{\omega_{m\mathbf{k}',l\mathbf{k}'}+2i\eta}-\frac{j^\alpha_{m\mathbf{k}',l\mathbf{k}}j^\mu_{l\mathbf{k},n\mathbf{k}}}{\omega_{l\mathbf{k},n\mathbf{k}}+2i\eta}\right)
    .
\end{equation}

\section{Density matrix perturbation theory}

One way of calculating nonlinear optical responses is to calculate the expected value of the current with respect to a perturbed density matrix at some order in perturbation theory. Specifically, we have
\begin{equation}
    \mathcal{H}(t) = \mathcal{H}^{(0)} + V(t)
\end{equation}
\begin{equation}
    \rho(t) = \rho^{(0)} + \rho^{(1)}(t) + \rho^{(2)}(t) + \hdots
\end{equation}
where $\rho^{(N)}(t)$ is at $N$-th order in $V(t)$. The goal is to compute the expected value of the current $\vec{J}$ with respect to a perturbed density matrix at some total frequency $\omega^{(N)} = \sum_{j=1}^N \omega_j$,
\begin{equation}
    \langle \vec{\tilde{J}}(\omega^{(N)}) \rangle = \text{tr}\left\{\tilde{\rho}^{(2)}(\omega^{(N)})\vec{J}\right\}
\end{equation}
with $\tilde{\rho}(\omega) = \int dt \,e^{i\omega t} \rho(t)$. Here we outline a procedure for doing that.

The density matrix evolves under the Hamiltonian via the equation of motion
\begin{equation}
    i\hbar \frac{\partial}{\partial t}\rho(t) = [\mathcal{H}(t),\rho(t)]
    .\label{eom}
\end{equation}
Writing this in the eigenbasis of the unperturbed Hamiltonian $\mathcal{H}^{(0)}|n\rangle = \epsilon_n|n\rangle$, we can write,
\begin{equation}
    \left(i\hbar\frac{\partial}{\partial t}-\epsilon_{mn}\right)\rho_{mn}(t) = [V(t),\rho(t)]_{mn}
\end{equation}
where $\epsilon_{mn} \equiv \epsilon_m - \epsilon_n$. This allows us to solve for $\rho^{(N)}(t)$ perturbatively in $V(t)$,
\begin{equation}
    \left(i\hbar\frac{\partial}{\partial t}-\epsilon_{mn}\right)\rho^{(N)}_{mn}(t) = [V(t),\rho^{(N-1)}(t)]_{mn}
    .
\end{equation}
In frequency domain, the products between $V(t)$ and $\rho^{(N-1)}(t)$ become a convolution over frequency,
\begin{equation}
    (\hbar\omega - \epsilon_{mn})\tilde{\rho}^{(N)}_{mn}(\omega) = \int d\omega'~[\tilde{V}(\omega-\omega'), \tilde{\rho}^{(N-1)}(\omega')]_{mn}
    .
\end{equation}
(We will hereafter drop the tildes, with the understanding that $V(\omega)$ and $\rho(\omega)$ refer to the frequency-domain functions when $\omega$ is the argument.) In the cases of interest to us, $V(\omega-\omega')$ is a discrete sum of pure frequencies,
\begin{equation}
    V(\omega-\omega') = \sum_p V(\omega_p) \delta(\omega-\omega'-\omega_p)
    .
\end{equation}
For example, when modeling optical responses to a plane wave electromagnetic field, we will assume $V(\omega-\omega') = V(\Omega) e^{-i\Omega t} \delta_{\omega-\omega'-\Omega} + V(-\Omega) e^{i\Omega t}\delta_{\omega-\omega'+\Omega}$. We therefore have in full generality
\begin{equation}
    \rho^{(N)}_{mn}(\omega^{(N)}) = \sum_{\{\omega_j\}_N}\frac{\left[V(\omega_N),\rho^{(N-1)}(\omega^{(N-1)})\right]_{mn}}{\hbar\omega^{(N)}-\epsilon_{mn}}
\end{equation}
where $\omega^{(N)}=\sum_{j=1}^N\omega_j$ and $\sum_{\{\omega_j\}_N}$ is a sum over all ways to take $N$ pure frequencies from $V(\omega-\omega')$ such that the total is $\omega^{(N)}$. As a technical matter, we will need to introduce an infinitesimal phenomenological relaxation rate $\eta \to 0^+$ as a positive imaginary shift to each frequency $\omega_j \to \omega_j + i\eta$. This can be introduced at the equation of motion level by adding a term $-\hbar\eta\frac{\partial}{\partial t}\rho(t)$ to the left side of Eq. (\ref{eom}). We therefore have the final result,
\begin{equation}
    \rho^{(N)}_{mn}(\omega^{(N)}) = \sum_{\{\omega_j\}_N}\frac{\left[V(\omega_N),\rho^{(N-1)}(\omega^{(N-1)})\right]_{mn}}{\hbar\omega^{(N)}-\epsilon_{mn}+iN\hbar\eta}
    .
\end{equation}
Note the factor of $N$ multiplying $\eta$.

Let us carry this out to first order. Assume $\rho^{(0)}_{mn} = \delta_{mn} f_n$, where $f_n$ is the Fermi occupation factor for state $n$. We then have, (absorbing a factor of $1/\hbar$ into the definition of $V$ and defining $\hbar\omega_{mn} = \epsilon_{mn}$)
\begin{equation}
    \rho^{(1)}_{mn}(\omega) = \frac{f_{nm}V_{mn}(\omega)}{\omega-\omega_{mn}+i\eta}
\end{equation}
Now we consider second order. Assuming the perturbation is at a single pure frequency $V(\omega-\omega') = V(\Omega) e^{-i\Omega t} \delta_{\omega-\omega'-\Omega} + V(-\Omega) e^{i\Omega t}\delta_{\omega-\omega'+\Omega}$, we can consider $\rho^{(2)}(\pm 2\Omega)$ or $\rho^{(2)}(0)$. For the purposes of photocurrent responses, we need $\rho^{(2)}(0)$, which has two contributions: $(\omega_1 = \omega,\omega_2=-\omega)$ and $(\omega_1=-\omega,\omega_2=\omega)$,
\begin{align}
    \rho^{(2)}_{lm}(0) &= \frac{1}{-\omega_{lm}+2i\eta}[V(\omega),\rho^{(1)}(-\omega)]_{lm} + [\omega \to -\omega] \\
    &= \frac{1}{\omega_{ml}+2i\eta}\sum_{n}\left(\frac{V_{ln}(\omega)V_{nm}(-\omega)f_{mn}}{\omega-\omega_{nm}+i\eta}-\frac{V_{ln}(-\omega)f_{nl}V_{nm}(\omega)}{\omega-\omega_{ln}+i\eta}\right) + [\omega\to -\omega].
\end{align}
Tracing this against the current operator (assuming the current operator is not itself perturbed by $V(t)$---not the case in velocity gauge light-matter coupling), we therefore have,
\begin{align}
    \langle \vec{J}(0) \rangle^{(2)} &= \text{tr}\left\{\rho^{(2)}(0)\vec{J}\right\} \\
    &= \sum_{nml} \left(\frac{V_{ln}(\omega)V_{nm}(-\omega)f_{mn}}{\omega-\omega_{nm}+i\eta} - \frac{V_{ln}(-\omega)f_{nl}V_{nm}(\omega)}{\omega-\omega_{ln}+i\eta}\right)\frac{\vec{j}_{ml}}{\omega_{ml}+2i\eta} + [\omega\to-\omega]
\end{align}
Relabeling indices $l \to n$, $n \to m$, $m\to l$ in the second term, we have
\begin{equation}
    \langle \vec{J}(0) \rangle^{(2)} = \sum_{nml} \frac{f_{mn}}{\omega+\omega_{mn}+i\eta}V_{nm}(-\omega)\left(\frac{\vec{j}_{ml}V_{ln}(\omega)}{\omega_{ml}+2i\eta} - \frac{V_{ml}(\omega)\vec{j}_{ln}}{\omega_{ln}+2i\eta}\right) + [\omega \to -\omega]
    .
    \label{J 2}
\end{equation}
Finally, we consider the states $n$, $m$, and $l$ to be indexed by band indices \textit{and} momentum, which can be different if $V(\omega)$ is promoted to $V(\omega,\mathbf{q})$. One can recover the diagrammatic results of the previous section of one substitutes,
\begin{equation}
    V_{mn}(\omega,\mathbf{q}) = \frac{e}{i\hbar\omega} E_\alpha(\omega,\mathbf{q})\langle \psi_{m\mathbf{k}'} | j^\alpha_{\mathbf{q}} | \psi_{n\mathbf{k}} \rangle.
\end{equation}
Importantly, Eq. (\ref{J 2}) corresponds \textit{only} to the part of the conductivity resulting from the first diagram in the diagrammatic analysis, i.e. the one with three separate one-photon vertices. One needs to add this to a response calculated from a first-order-perturbed density matrix traced with the \textit{diamagnetic} current operator $j^{\mu\alpha}_{\mathbf{0},\mathbf{q}}$ in order to obtain the correct response at finite frequency. The diagrams have the advantage of pictorially keeping track of all the required terms to obtain a regularized response \cite{parker2019, gassner2023}.

\section{Shift and Injection currents}

Summing Diagrams 1 and 2 is sufficient to obtain a regularized photovoltaic response for frequencies away from zero, i.e. they are the only diagrams that contribute at resonance between states that differ in energy (momentum) by $\hbar\omega$ ($\hbar\mathbf{q}$). Summing their matrix element parts $\mathcal{M}^{\mu\alpha\beta}_{m\mathbf{k}',n\mathbf{k}}=\left[\mathcal{M}^{\mu\alpha\beta}_{m\mathbf{k}',n\mathbf{k}}\right]_{(1)}+\left[\mathcal{M}^{\mu\alpha\beta}_{m\mathbf{k}',n\mathbf{k}}\right]_{(2)}$, we find
\begin{equation}
    \sigma^{\mu\alpha\beta}_\text{PV}(\omega,\mathbf{q}) \equiv \sigma^{\mu\alpha\beta}(0;q,-q) = \frac{ie^3}{\hbar^2\omega^2}\frac{1}{L^{2d}} \sum_{\mathbf{k},\mathbf{k}'}\delta_{\mathbf{k}'-\mathbf{k}-\mathbf{q}}\sum_{nm}f_{m\mathbf{k},n\mathbf{k}'}\left(\frac{\mathcal{M}^{\mu\alpha\beta}_{m\mathbf{k}',n\mathbf{k}}}{\omega_{m\mathbf{k}',n\mathbf{k}}-\omega+i\eta}-\frac{\left[\mathcal{M}^{\mu\beta\alpha}_{m\mathbf{k}',n\mathbf{k}}\right]^*}{\omega_{m\mathbf{k}',n\mathbf{k}}-\omega-i\eta}\right)
    \label{sigma PV}
\end{equation}
\begin{equation}
    \mathcal{M}^{\mu\alpha\beta}_{m\mathbf{k}',n\mathbf{k}} = ij^\beta_{n\mathbf{k},m\mathbf{k}'}\left[j^{\mu\alpha}_{m\mathbf{k}',n\mathbf{k}}+\sum_l\left(\frac{j^\mu_{m\mathbf{k}',l\mathbf{k}'}j^\alpha_{l\mathbf{k}',n\mathbf{k}}}{\omega_{m\mathbf{k}',l\mathbf{k}'}+2i\eta}-\frac{j^\alpha_{m\mathbf{k}',l\mathbf{k}}j^\mu_{l\mathbf{k},n\mathbf{k}}}{\omega_{l\mathbf{k},n\mathbf{k}}+2i\eta}\right)\right]
    .
\end{equation}
This is conventionally separated into two contributions: Regarding $\eta \equiv \frac{1}{2}\tau^{-1}$ as a finite scattering rate, there is a linear-in-$\tau$ contribution that comes from the ``band-diagonal'' terms in the sum over $l$ where the denominator nearly vanishes. This is referred to as the ``injection current'' contribution,
\begin{align}
    \left[\mathcal{M}^{\mu\alpha\beta}_{m\mathbf{k}',n\mathbf{k}}\right]_\text{inj.} &= \tau j^\beta_{n\mathbf{k},m\mathbf{k}'}\left(j^\mu_{m\mathbf{k}',m\mathbf{k}'}-j^\mu_{n\mathbf{k},n\mathbf{k}}\right)j^\alpha_{m\mathbf{k}',n\mathbf{k}} \\
    &\equiv \tau j^\beta_{n\mathbf{k},m\mathbf{k}'}j^\alpha_{m\mathbf{k}',n\mathbf{k}} \left(v^\mu_{m\mathbf{k}'}-v^\mu_{n\mathbf{k}}\right)
\end{align}
where $v^\mu_{n\mathbf{k}} \equiv \partial_k\omega_{n\mathbf{k}} \equiv \frac{1}{\hbar}\langle u_{n\mathbf{k}} | \partial_{k_\mu} h_\mathbf{k} | u_{n\mathbf{k}}\rangle$ is the intraband velocity of band $n$ at momentum $\mathbf{k}$.

The remaining piece is called the ``shift current'' contribution. Assuming $\eta$ to be much smaller than the smallest gap between bands of distinct energy, and using
\begin{equation}
    j^\mu_{l\mathbf{k},n\mathbf{k}} \equiv v^\mu_{l\mathbf{k},n\mathbf{k}} \equiv \frac{1}{\hbar}\langle u_{l\mathbf{k}} | \nabla^\mu_\mathbf{k}h_\mathbf{k}|u_{n\mathbf{k}}\rangle \overset{l\ne n}{=} -\omega_{l\mathbf{k},m\mathbf{k}}\langle u_{l\mathbf{k}} | \nabla^\mu_\mathbf{k} u_{n\mathbf{k}}\rangle
\end{equation}
where we abbreviate $\nabla^\mu_\mathbf{k}\equiv \partial/\partial k_\mu$, we find
\begin{align}
    \left[\mathcal{M}^{\mu\alpha\beta}_{m\mathbf{k}',n\mathbf{k}}\right]_\text{shift} &= ij^\beta_{n\mathbf{k},m\mathbf{k}'}\left[j^{\mu\alpha}_{m\mathbf{k}',n\mathbf{k}}+\sum_{l\ne m}\frac{j^\mu_{m\mathbf{k}',l\mathbf{k}'}j^\alpha_{l\mathbf{k}',n\mathbf{k}}}{\omega_{m\mathbf{k}',l\mathbf{k}'}}-\sum_{l\ne n}\frac{j^\alpha_{m\mathbf{k}',l\mathbf{k}}j^\mu_{l\mathbf{k},n\mathbf{k}}}{\omega_{l\mathbf{k},n\mathbf{k}}}\right] \\
    &= ij^\beta_{n\mathbf{k},m\mathbf{k}'}\left[j^{\mu\alpha}_{m\mathbf{k}',n\mathbf{k}}-\sum_{l\ne m}\langle u_{m\mathbf{k}'}|\nabla^\mu_{\mathbf{k}'} u_{l\mathbf{k}'}\rangle j^\alpha_{l\mathbf{k}',n\mathbf{k}}+\sum_{l\ne n}j^\alpha_{m\mathbf{k}',l\mathbf{k}}\langle u_{l\mathbf{k}}|\nabla^\mu_{\mathbf{k}} u_{n\mathbf{k}}\rangle \right] \\
    &= ij^\beta_{n\mathbf{k},m\mathbf{k}'}\left[j^{\mu\alpha}_{m\mathbf{k}',n\mathbf{k}}+\sum_{l\ne m}\langle \nabla^\mu_{\mathbf{k}'} u_{m\mathbf{k}'}| u_{l\mathbf{k}'}\rangle j^\alpha_{l\mathbf{k}',n\mathbf{k}}+\sum_{l\ne n}j^\alpha_{m\mathbf{k}',l\mathbf{k}}\langle u_{l\mathbf{k}}|\nabla^\mu_{\mathbf{k}} u_{n\mathbf{k}}\rangle \right] \\
    &= ij^\beta_{n\mathbf{k},m\mathbf{k}'}\left[j^{\mu\alpha}_{m\mathbf{k}',n\mathbf{k}}+(\nabla^\mu_{\mathbf{k}} + \nabla^\mu_{\mathbf{k}'} - i\mathcal{A}^\mu_{m\mathbf{k}'}+i\mathcal{A}^\mu_{n\mathbf{k}})j^\alpha_{n\mathbf{k},m\mathbf{k}'} - \langle u_{m\mathbf{k}'}|\left(\nabla^\mu_{\mathbf{k}}+\nabla^\mu_{\mathbf{k}'}\right)\left(e^{-i\mathbf{k}'\cdot\mathbf{r}}j^\alpha _\mathbf{q}e^{i\mathbf{k}\cdot\mathbf{r}}\right)|u_{n\mathbf{k}}\rangle\right] \\
    &= j^\beta_{n\mathbf{k},m\mathbf{k}'}\left(i\nabla^\mu_\mathbf{k}+i\nabla^\mu_{\mathbf{k}'}+\mathcal{A}^\mu_{m\mathbf{k}'}-\mathcal{A}^\mu_{n\mathbf{k}}\right)j^\alpha_{m\mathbf{k}',n\mathbf{k}}-j^\beta_{n\mathbf{k},m\mathbf{k}'}\langle \psi_{m\mathbf{k}'} | \left(j^{\mu\alpha}_{\mathbf{0},\mathbf{q}} - i[j^\alpha_\mathbf{q},r^\mu]\right) | \psi_{n\mathbf{k}}\rangle
    \label{second term}
\end{align}
where $\mathcal{A}^\mu_{n\mathbf{k}} \equiv i\langle u_{n\mathbf{k}}|\nabla^\mu_\mathbf{k} u_{n\mathbf{k}}\rangle$ is the intraband (Abelian) Berry connection, and $r^\mu$ is the position operator. The second term vanishes in the $\mathbf{q}\to 0$ limit, in which $e^{-i\mathbf{k}\cdot\mathbf{r}}j^{\mu\alpha}_{\mathbf{0},\mathbf{0}}e^{i\mathbf{k}\cdot\mathbf{r}} = \nabla^\mu_\mathbf{k}\nabla^\alpha_\mathbf{k}h_\mathbf{k} = [-ir^\mu,\nabla^\alpha_\mathbf{k}h_\mathbf{k}]\equiv e^{-i\mathbf{k}\cdot\mathbf{r}}[-ir^\mu,j^\alpha_\mathbf{0}]e^{i\mathbf{k}\cdot\mathbf{r}}$. The first term then recovers the conventional shift current \cite{sipe2000}. For finite $\mathbf{q}$, one can show from a Ward identity \cite{mckay2024} that
\begin{align}
    q'_\mu j^{\mu\alpha}_{\mathbf{q}',\mathbf{q}} &= [j^\alpha_\mathbf{q},\rho_{\mathbf{q}'}] \\
    &= j^\alpha_{\mathbf{q}+\mathbf{q}'} - j^\alpha_\mathbf{q} \\
    &= e^{-i\mathbf{q}'\cdot\mathbf{r}}j^\alpha_\mathbf{q}e^{i\mathbf{q}'\cdot\mathbf{r}} - j^\alpha_{\mathbf{q}} \\
    &= iq'_\mu[j^\alpha_\mathbf{q},r^\mu] + \mathcal{O}(\mathbf{q}'^2)
\end{align}
so provided that $j^{\mu\alpha}_{\mathbf{q}',\mathbf{q}}$ is differentiable in $\mathbf{q}'$ near $\mathbf{q}'=0$, the second term in Eq. (\ref{second term}) vanishes. Assuming this to be the case, we have
\begin{equation}
    \left[\mathcal{M}^{\mu\alpha\beta}_{m\mathbf{k}',n\mathbf{k}}\right]_\text{shift} = j^\beta_{n\mathbf{k},m\mathbf{k}'}(i\nabla^\mu_\mathbf{k}+i\nabla^\mu_{\mathbf{k}'}+\mathcal{A}^\mu_{m\mathbf{k}'}-\mathcal{A}^\mu_{n\mathbf{k}})j^\alpha_{m\mathbf{k}',n\mathbf{k}}
    .
\end{equation}
Explicitly writing the conductivity for the shift and injection currents, we can take the $\eta\to 0$ limit in Eq. (\ref{sigma PV}) (but still keeping the ``intraband" scattering rate $\tau^{-1}$ finite in the injection current case) to obtain delta-function forms for the conductivities. Using $\lim_{\eta\to 0^+}\frac{1}{x-i\eta} = i\pi\delta(x) + \mathcal{P}(1/x)$, where $\mathcal{P}$ denotes the Cauchy principal value,
\begin{align}
    \begin{split}
    \sigma^{\mu\alpha\beta}_\text{PV}(\omega,\mathbf{q}) = \frac{-\pi e^3}{\hbar^2\omega^2}\frac{1}{L^{2d}}\sum_{\mathbf{k},\mathbf{k}'}\delta_{\mathbf{k}'-\mathbf{k}-\mathbf{q}}\sum_{mn}f_{m\mathbf{k}',n\mathbf{k}} \bigg[&
        \delta(\omega_{m\mathbf{k}',n\mathbf{k}}-\omega)\left(\mathcal{M}^{\mu\alpha\beta}_{m\mathbf{k}',n\mathbf{k}}+\left[\mathcal{M}^{\mu\beta\alpha}_{m\mathbf{k}',n\mathbf{k}}\right]^*\right) \\
        &+ \mathcal{P}\left(\frac{1}{\omega_{m\mathbf{k}',n\mathbf{k}}-\omega}\right)\left(\mathcal{M}^{\mu\alpha\beta}_{m\mathbf{k}',n\mathbf{k}}-\left[\mathcal{M}^{\mu\beta\alpha}_{m\mathbf{k}',n\mathbf{k}}\right]^*\right)\bigg].
    \end{split}
\end{align}
For the injection current, it is easy to see that $\mathcal{M}^{\mu\alpha\beta}_{m\mathbf{k}',n\mathbf{k}}-\left[\mathcal{M}^{\mu\beta\alpha}_{m\mathbf{k}',n\mathbf{k}}\right]^*$ vanishes at every $\mathbf{k}$, $\mathbf{k}'$. For the shift current, it generically does not:
\begin{align}
    \begin{split}
        \left[\mathcal{M}^{\mu\alpha\beta}_{m\mathbf{k}',n\mathbf{k}}\right]_\text{shift} - \left[\mathcal{M}^{\mu\alpha\beta}_{m\mathbf{k}',n\mathbf{k}}\right]_\text{shift}^* &= j^\beta_{n\mathbf{k},m\mathbf{k}'}(i\nabla^\mu_\mathbf{k}+i\nabla^\mu_{\mathbf{k}'}+\mathcal{A}^\mu_{m\mathbf{k}'}-\mathcal{A}^\mu_{n\mathbf{k}})j^\alpha_{m\mathbf{k}',n\mathbf{k}} \\
        &~~~-j^\alpha_{m\mathbf{k}',n\mathbf{k}}(-i\nabla^\mu_\mathbf{k}-i\nabla^\mu_{\mathbf{k}'}+\mathcal{A}^\mu_{m\mathbf{k}'}-\mathcal{A}^\mu_{n\mathbf{k}})j^\beta_{n\mathbf{k},m\mathbf{k}'}
    \end{split}
    \\
    &= (i\nabla^\mu_\mathbf{k}+i\nabla^\mu_{\mathbf{k}'})\left(j^\beta_{n\mathbf{k},m\mathbf{k}'}j^\alpha_{m\mathbf{k}',n\mathbf{k}}\right).
\end{align}
Care must be taken in ensuring the principal part vanishes upon $\mathbf{k}$-integration. This is discussed in the $\mathbf{q}\to 0$ limit in the context of sum rules \cite{aversa1995} and in a recent review \cite{dai2023recent}. We at last obtain,
\begin{align}
    \begin{split}
    \sigma^{\mu\alpha\beta}_\text{PV}(\omega,\mathbf{q}) = \frac{-\pi e^3}{\hbar^2\omega^2}\frac{1}{L^{2d}}\sum_{\mathbf{k},\mathbf{k}'}\delta_{\mathbf{k}'-\mathbf{k}-\mathbf{q}}\sum_{mn}f_{m\mathbf{k}',n\mathbf{k}}\,
        \delta(\omega_{m\mathbf{k}',n\mathbf{k}}-\omega)\left(\mathcal{M}^{\mu\alpha\beta}_{m\mathbf{k}',n\mathbf{k}}+\left[\mathcal{M}^{\mu\beta\alpha}_{m\mathbf{k}',n\mathbf{k}}\right]^*\right)
    \end{split}
    \label{sigma PV delta form}
\end{align}
\begin{equation}
    \mathcal{M}^{\mu\alpha\beta}_{m\mathbf{k}',n\mathbf{k}} = j^\beta_{n\mathbf{k},m\mathbf{k}'}\left(i\nabla^\mu_{\mathbf{k}'}+i\nabla^\mu_\mathbf{k}+\mathcal{A}^\mu_{m\mathbf{k}'}-\mathcal{A}^\mu_{n\mathbf{k}} + \tau v^\mu_{m\mathbf{k}'}-\tau v^\mu_{n\mathbf{k}}\right)j^\alpha_{m\mathbf{k}',n\mathbf{k}}.
\end{equation}
Note that the apparent $1/\omega^2$ divergence is taken care of by the delta function $\delta(\omega_{m\mathbf{k}',n\mathbf{k}}-\omega)$. Making the substitution $\omega\to\omega_{m\mathbf{k}',n\mathbf{k}}$ is crucial for recovering ``length gauge" expressions for the response function, as well as obtaining the multipole forms that we describe in the following sections.

\subsection{The $\mathbf{q}\to 0$ limit}

As a quick check, we verify that the conventional shift and injection currents are recovered in the $\mathbf{q}\to 0$ limit. Using the delta function to replace $\omega\to\omega_{m\mathbf{k}',n\mathbf{k}}$ and setting $\mathbf{k}'=\mathbf{k}$,
\begin{equation}
    \frac{\mathcal{M}^{\mu\alpha\beta}_{m\mathbf{k},n\mathbf{k}}}{(\omega_{m\mathbf{k},n\mathbf{k}})^2} = r^\beta_{nm\mathbf{k}}\left(i\nabla^\mu_\mathbf{k}+\mathcal{A}^\mu_{m\mathbf{k}}-\mathcal{A}^\nu_{m\mathbf{k}}+\tau v^\mu_{m\mathbf{k}}-\tau v^\mu_{n\mathbf{k}}\right)r^\alpha_{mn\mathbf{k}}
\end{equation}
where we use the fact that $j^\alpha_{m\mathbf{k},n\mathbf{k}} \equiv v^\mu_{mn\mathbf{k}} = i\omega_{mn\mathbf{k}}r^\mu_{mn\mathbf{k}}$ for $m\ne n$. The injection current is the part proportional to the scattering time $\tau$, which is phenomenological at this level and is sometimes defined in the literature with a relative factor of $2$. The shift current is the remaining piece, and is oftentimes denoted concisely as $i r^\beta_{nm\mathbf{k}}r^{\alpha;\mu}_{mn\mathbf{k}}$ where $\mathcal{O}^{;\mu}_{mn\mathbf{k}} \equiv (\nabla^\mu_\mathbf{k} - i(\mathcal{A}^\mu_{m\mathbf{k}}-\mathcal{A}^\mu_{n\mathbf{k}}))\mathcal{O}_{mn\mathbf{k}}$ is shorthand for the $U(1)^2$ gauge-covariant derivative. This is also often expressed in terms of the so-called shift vector $\mathcal{R}^{\mu\alpha}_{mn}$ as follows, \cite{cook2017,ahn2020}
\begin{equation}
    ir^\beta_{nm\mathbf{k}}r^{\alpha;\mu}_{mn\mathbf{k}} - ir^\alpha_{mn\mathbf{k}}r^{\beta;\mu}_{nm\mathbf{k}} = r^\beta_{nm}r^\alpha_{mn}\left(\mathcal{R}^{\mu\alpha}_{mn\mathbf{k}}-\mathcal{R}^{\mu\beta}_{nm\mathbf{k}}\right)
\end{equation}
\begin{equation}
    \mathcal{R}^{\mu\alpha}_{mn\mathbf{k}} = \mathcal{A}^\mu_{m\mathbf{k}}-\mathcal{A}^\mu_{n\mathbf{k}} + i\nabla^\mu_\mathbf{k}\log r^\alpha_{mn\mathbf{k}}
    .
\end{equation}

\subsection{Expansion to first order in $\mathbf{q}$}

We now expand the photovoltaic conductivity (\ref{sigma PV delta form}) to first order in the incident wavevector $\mathbf{q}$. The crucial piece for our purposes is the expansion of the current operator matrix elements $j^\mu_{m\mathbf{k}',n\mathbf{k}} \equiv \langle \psi_{m\mathbf{k}'} | j^\mu_{\mathbf{k}'-\mathbf{k}} | \psi_{n\mathbf{k}}\rangle $, noting that the momentum delta function in our response function enforces $\mathbf{k}'=\mathbf{k}+\mathbf{q}$. It is simplest to adopt the parameterization $\mathbf{k}'\to\mathbf{k}+\frac{\mathbf{q}}{2}$, $\mathbf{k}\to\mathbf{k}-\frac{\mathbf{q}}{2}$. Remembering also the delta function in frequency, which turns $1/\omega^2 = 1/(\omega_{m\mathbf{k}',n\mathbf{k}})^2$, the natural object to expand is
\begin{equation}
    \frac{j^\mu_{(m,\mathbf{k}+\frac{\mathbf{q}}{2}),(n,\mathbf{k}-\frac{\mathbf{q}}{2})}}{i\omega_{(m,\mathbf{k}+\frac{\mathbf{q}}{2}),(n,\mathbf{k}-\frac{\mathbf{q}}{2})}} = \frac{v^\mu_{mn\mathbf{k}}}{i\omega_{mn\mathbf{k}}} + \frac{q_\nu}{2}\left(\frac{i\left\{\mathcal{A}^\nu,v^\mu\right\}_{mn\mathbf{k}}}{i\omega_{mn\mathbf{k}}}-\frac{v^\mu_{mn\mathbf{k}}}{i(\omega_{mn\mathbf{k}})^2}(v^\nu_{mm\mathbf{k}}+v^\nu_{nn\mathbf{k}})\right) + \mathcal{O}(\mathbf{q}^2)
    ,\label{current operator expansion}
\end{equation}
where $v^\mu_{mn} = \langle u_{m\mathbf{k}} | v^\mu_\mathbf{k} | u_{n\mathbf{k}} \rangle$ denotes the Bloch velocity operator matrix elements and $\mathcal{A}^\mu_{mn} = i\langle u_{m\mathbf{k}} | \nabla^\mu_\mathbf{k} u_{n\mathbf{k}}\rangle$ denotes the non-Abelian Berry connection. Here, we used two facts,
\begin{equation}
\begin{split}
    e^{-i(\mathbf{k}+\frac{\mathbf{q}}{2})\cdot\mathbf{r}} j^\mu_\mathbf{q} e^{i(\mathbf{k}-\frac{\mathbf{q}}{2})\cdot\mathbf{r}} &= e^{-i\mathbf{k}\cdot\mathbf{r}}v^\mu e^{i\mathbf{k}\cdot\mathbf{r}} + \mathcal{O}(\mathbf{q}^2) \\
    &\equiv v^\mu_\mathbf{k} + \mathcal{O}(\mathbf{q}^2)
    \label{current operator assumption}
\end{split}
\end{equation}
\begin{align}
    \left.\nabla^\nu_\mathbf{q}\langle u_{m,\mathbf{k}+\frac{\mathbf{q}}{2}} | v^\mu_\mathbf{k} | u_{n,\mathbf{k}-\frac{\mathbf{q}}{2}} \rangle\right|_{\mathbf{q}\to 0} &= \frac{1}{2}\left(\langle \partial^\nu_\mathbf{k} u_{m\mathbf{k}}| v^\mu_\mathbf{k} | u_{n\mathbf{k}}\rangle - \langle u_{m\mathbf{k}}| v^\mu_\mathbf{k} | \partial^\nu_\mathbf{k}u_{n\mathbf{k}}\rangle\right) \\
    &= \frac{i}{2}\sum_l \left(\mathcal{A}^\nu_{ml}v^\mu_{ln} + v^\mu_{ml}\mathcal{A}^\nu_{ln}\right) \equiv \frac{i}{2}\left\{\mathcal{A}^\nu,v^\mu\right\}_{mn}
\end{align}
where we hereafter suppress the $\mathbf{k}$-dependence to declutter notation. The anticommutator $\frac{i}{2}\left\{\mathcal{A}^\nu,v^\mu\right\}_{mn}$ has been called $(B^{(\text{orb.})})_{mn,\mu\nu}$ in Ref. \cite{malashevich2010}, where $\text{(orb.)}$ distinguishes it from the spin contribution $B^{(\text{spin})}_{mn,\mu\nu}=-\frac{i}{m_\text{e}}\epsilon_{\mu\nu\rho}\langle u_{m}|S^\rho|u_{n}\rangle$ to the magnetic dipole current (which we neglect here). We emphasize that Eq. (\ref{current operator assumption}) is strictly speaking an assumption. It is valid for Hamiltonians of the form $\mathcal{H}_0 = \mathbf{p}^2/2m + V(\mathbf{r})$, where the corresponding Bloch Hamiltonian $h_\mathbf{k} = e^{-i\mathbf{k}\cdot\mathbf{r}}\mathcal{H}_0 e^{i\mathbf{k}\cdot\mathbf{r}}$ satisfies $\nabla^\mu_\mathbf{k} h_\mathbf{k} \equiv v^\mu_\mathbf{k} = e^{-i\mathbf{k}\cdot\mathbf{r}}\frac{\mathbf{p}}{m} e^{i\mathbf{k}\cdot\mathbf{r}}$. But one should use caution when making the assumption in Eq. (\ref{current operator assumption}) in models with a highly truncated band basis (e.g. tight binding models). One consequence, as pointed out by Ref. \cite{mckay2024}, is that the resulting current operators may not obey charge conservation / gauge invariance. Deviations from the approximation $\langle \psi_{m,\mathbf{k}+\frac{\mathbf{q}}{2}} | j^\mu_\mathbf{q} | \psi_{n,\mathbf{k}-\frac{\mathbf{q}}{2}}\rangle \approx \langle u_{m,\mathbf{k}+\frac{\mathbf{q}}{2}} | v^\mu_\mathbf{k} | u_{n,\mathbf{k}-\frac{\mathbf{q}}{2}}\rangle$ are of fundamental interest to the study of finite-wavevector response functions, and will be the subject of future work.

Proceeding under this assumption, we take inspiration from Ref. \cite{ocana2023} and separate out the band-diagonal parts of the anticommutator $\left\{\mathcal{A}^\nu,v^\mu\right\}_{mn}$,
\begin{equation}
    \left\{\mathcal{A}^\nu,v^\mu\right\}_{mn} = \sum_{l\ne m,n}\left(\mathcal{A}^\nu_{ml} v^\mu_{ln} + v^\mu_{ml}\mathcal{A}^\nu_{ln}\right) + (\mathcal{A}^\nu_{mm}+\mathcal{A}^\nu_{nn})v^\mu_{mn} + \mathcal{A}^\nu_{mn}(v^\mu_{mm}+v^\mu_{nn})
    .
    \label{origin dependence separation}
\end{equation}
The physical interpretation of the second term, when integrated over $\mathbf{k}$, is an average Wannier center between bands $m$ and $n$ times an interband velocity. As discussed in Ref. \cite{ocana2023}, this reflects the origin dependence of the multipole currents. Occasionally this term will not contribute to a particular optical response function (such as the 1D photocurrent we consider in the main text), but not always. For example, we will find that it contributes to the ``sum of Berry curvatures" term in the linear-in-$\mathbf{q}$ photocurrent first noted by \cite{xie2025}. Nevertheless, separating out the origin-independent multipole portion of the response is helpful for separating out the part of the response that depends on ``extrinsic" quantum geometry, as we will discuss.

It is conceptually helpful to separate the linear-in-$\mathbf{q}$ correction into symmetric and antisymmetric parts with respect to $\mu\leftrightarrow\nu$, which correspond respectively to the electric quadrupole and magnetic dipole contributions.
\begin{equation}
    \frac{j^\mu_{(m,\mathbf{k}+\frac{\mathbf{q}}{2}),(n,\mathbf{k}-\frac{\mathbf{q}}{2})}}{i\omega_{(m,\mathbf{k}+\frac{\mathbf{q}}{2}),(n,\mathbf{k}-\frac{\mathbf{q}}{2})}} \equiv r^\mu_{mn} + q_\nu\left(\frac{i}{2}Q^{\nu\mu}_{mn} + \frac{\varepsilon^{\nu\mu\rho}}{\omega_{mn}}M^\rho_{mn}\right) + \mathcal{O}(\mathbf{q}^2)
\end{equation}
where
\begin{align}
    Q^{\nu\mu}_{mn} &= \frac{1}{2}\frac{\sum_{l\ne m,n}(\mathcal{A}^\nu_{ml}v^\mu_{ln}+v^\mu_{ml}\mathcal{A}^\nu_{ln})}{i\omega_{mn}} + \frac{\mathcal{A}^\nu_{mm}+\mathcal{A}^\nu_{nn}}{2}\frac{v^\mu_{mn}}{i\omega_{mn}} + [\mu\leftrightarrow\nu] \\
    &= \frac{1}{2}\frac{\sum_{l\ne m,n}\left\{r^\nu_{ml}(i\omega_{ln}r^\mu_{ln})+(i\omega_{ml}r^\mu_{ml})r^\nu_{ln} + [\mu\leftrightarrow\nu]\right\}}{i\omega_{mn}} + \frac{\mathcal{A}^\nu_{mm}+\mathcal{A}^\nu_{nn}}{2}r^\mu_{mn} + \frac{\mathcal{A}^\mu_{mm}+\mathcal{A}^\mu_{nn}}{2}r^\nu_{mn} \\
    &= \frac{\sum_{l\ne m,n}i(\omega_{ml}+\omega_{ln})\left(r^\nu_{ml}r^\mu_{ln}+r^\mu_{ml}r^\nu_{ln}\right)}{i\omega_{mn}} + \frac{\mathcal{A}^\nu_{mm}+\mathcal{A}^\nu_{nn}}{2}r^\mu_{mn} + \frac{\mathcal{A}^\mu_{mm}+\mathcal{A}^\mu_{nn}}{2}r^\nu_{mn} \\
    &= \sum_{l\ne m,n}\left(r^\nu_{ml}r^\mu_{ln}+r^\mu_{ml}r^\nu_{ln}\right) + \frac{\mathcal{A}^\nu_{mm}+\mathcal{A}^\nu_{nn}}{2}r^\mu_{mn} + \frac{\mathcal{A}^\mu_{mm}+\mathcal{A}^\mu_{nn}}{2}r^\nu_{mn} \\
    &\equiv \widetilde{Q}^{\nu\mu}_{mn} + \frac{\mathcal{A}^\nu_{mm}+\mathcal{A}^\nu_{nn}}{2}r^\mu_{mn} + \frac{\mathcal{A}^\mu_{mm}+\mathcal{A}^\mu_{nn}}{2}r^\nu_{mn}
\end{align}
and
\begin{align}
    \varepsilon^{\nu\mu\rho}M^\rho_{mn} &= -\frac{i}{4}\sum_{l\ne m,n}(\mathcal{A}^\nu_{ml}v^\mu_{ln}+v^\mu_{ml}\mathcal{A}^\nu_{ln}) - \frac{i}{4}(\mathcal{A}^\nu_{mm}+\mathcal{A}^\nu_{nn})v^\mu_{mn} - \frac{i}{2}\mathcal{A}^\nu_{mn}(v^\mu_{mm}+v^\mu_{nn}) - [\mu\leftrightarrow\nu] \\
    &\equiv \varepsilon^{\nu\mu\rho}\widetilde{M}^\rho_{mn} -\frac{i}{4}\left(\mathcal{A}^\nu_{mm}+\mathcal{A}^\nu_{nn}\right)v^\mu_{mn} + \frac{i}{4}v^\nu_{mn}\left(\mathcal{A}^\mu_{mm}+\mathcal{A}^\mu_{nn}\right)
    ,
\end{align}
where we use the fact that $v^\nu_{mn} = i\omega_{mn}r^\nu_{mn}$ and $r^\nu_{mn} = \mathcal{A}^\nu_{mn}$ for $m\ne n$ (we hereafter always assume $m\ne n$). Two nontrivial features of this multipole separation should be emphasized: (i) the last term in Eq. (\ref{current operator expansion}) (which came from expanding the denominator on the left-hand side) combined with the last term of Eq. (\ref{origin dependence separation}) to create a purely antisymmetric term, therefore only entering in the magnetic dipole part; and (ii) writing the restricted sum over $l$ in terms of the band energy differences leads to a term $\omega_{ml}+\omega_{ln}$ for the quadrupolar piece that precisely cancels the denominator $\omega_{mn}$, leading to an expression that appealingly only depends on the non-Abelian Berry connection.

Lastly, it will be useful later on to use the ``origin-independent" multipoles $\widetilde{Q}^{\nu\mu}_{mn}$ and $\widetilde{M}^\rho_{mn}$ defined in the last lines. For convenience, they are
\begin{equation}
    \widetilde{Q}^{\nu\mu}_{mn} = \frac{1}{2}\sum_{l\ne m,n} (r^\nu_{ml}r^\nu_{ln} + r^\mu_{ml}r^\nu_{ln})
\end{equation}
\begin{equation}
    \epsilon^{\nu\mu\rho}\widetilde{M}^\rho_{mn} = -\frac{i}{4}\sum_{l\ne m,n}(r^\nu_{ml}v^\mu_{ln}+v^\mu_{ml}r^\nu_{ln}-r^\mu_{ml}v^\nu_{ln}-v^\nu_{ml}r^\mu_{ln}) - \frac{i}{2}r^\nu_{mn}(v^\mu_{mm}+v^\mu_{nn}) + \frac{i}{2}(v^\nu_{mm}+v^\nu_{nn})r^\mu_{mn}
\end{equation}
Concretely, they are the parts of the multipoles that do not depend on the average intraband Berry connection $\frac{1}{2}\left(\mathcal{A}^\mu_{mm}+\mathcal{A}^\mu_{nn}\right)$. Note that our definition of the origin-independent magnetic dipole differs from the definition in Ref. \cite{ocana2023}; we include the part that is band-diagonal in velocity, since this is origin-independent provided that $m\ne n$. We can therefore alternatively write the expansion of the current operator as
\begin{equation}
    \left.\nabla^\nu_\mathbf{q}\frac{j^\mu_{(m,\mathbf{k}+\frac{\mathbf{q}}{2}),(n,\mathbf{k}-\frac{\mathbf{q}}{2})}}{i\omega_{(m,\mathbf{k}+\frac{\mathbf{q}}{2}),(n,\mathbf{k}-\frac{\mathbf{q}}{2})}}\right|_{\mathbf{q}\to 0} = \frac{i}{2}\widetilde{Q}^{\nu\mu}_{mn} + \frac{\varepsilon^{\nu\mu\rho}}{\omega_{mn}}\widetilde{M}^\rho_{mn} + i\frac{\mathcal{A}^\nu_{mm}+\mathcal{A}^\nu_{nn}}{2}r^\mu_{mn}
    .
\end{equation}
where the last term (the origin-dependent term) has both symmetric and antisymmetric parts with respect to $\mu\leftrightarrow\nu$.

We note that $M^\rho$ is defined such that its corresponding linear-in-$\mathbf{q}$ correction in terms of the magnetic field $\mathbf{B}$ is $B_\rho M^\rho_{mn}$ when $\omega = \omega_{mn}$. Recall from Ampere's Law,
\begin{equation}
\begin{split}
    \boldsymbol{\nabla}\times\mathbf{E} = -\frac{\partial}{\partial t}\mathbf{B} ~~~&\implies~~~iq_\nu \varepsilon^{\nu\mu\rho} E_\mu(\omega,\mathbf{q}) = i\omega B^\rho(\omega,\mathbf{q}) \\
    &\implies~~~B^\rho(\omega,\mathbf{q}) = q_\nu\frac{\varepsilon^{\nu\mu\rho}}{\omega}E_\mu(\omega,\mathbf{q})
    .
\end{split}
\end{equation}

With the expanded current operator matrix elements in hand, we can now expand the photovoltaic conductivity (\ref{sigma PV delta form}) to first order in the wavevector $\mathbf{q}$. We will demonstrate the derivation for the shift current only, as the injection current derivation can be carried out analogously. It is convenient to shift variables $\mathbf{k} \to \mathbf{k}-\frac{\mathbf{q}}{2}$ and sum over $\mathbf{k}'$ such that the delta function replaces $\mathbf{k}' \to \mathbf{k}'+\frac{\mathbf{q}}{2}$. Taking a $\mathbf{q}$-derivative then gives us the following form,
\begin{equation}
\begin{split}
    \left.\nabla_\mathbf{q}^\nu\sigma^{\mu\alpha\beta}_\text{PV}(\omega,\mathbf{q})\right|_{\mathbf{q}\to 0} = \frac{-\pi e^3}{\hbar^2\omega^2}\frac{1}{L^{2d}}\sum_{\mathbf{k},mn}\bigg(&f_{mn\mathbf{k}}\,
        \delta(\omega_{mn\mathbf{k}}-\omega)\left(\mathcal{M}^{\nu\mu\alpha\beta}_{mn\mathbf{k}}+\left[\mathcal{M}^{\nu\mu\beta\alpha}_{mn\mathbf{k}}\right]^*\right) \\
        &+ f_{mn\mathbf{k}}\delta'(\omega_{mn\mathbf{k}}-\omega)\frac{v^\mu_{m\mathbf{k}}+v^\mu_{n\mathbf{k}}}{2}\left(\mathcal{M}^{\mu\alpha\beta}_{mn\mathbf{k}}+\left[\mathcal{M}^{\mu\beta\alpha}_{mn\mathbf{k}}\right]^*\right) \\
        &+ \delta_{mn}f'_{n\mathbf{k}}\delta(\omega)\left(\mathcal{M}^{\mu\alpha\beta}_{nn\mathbf{k}} + \left[\mathcal{M}^{\mu\beta\alpha}_{nn\mathbf{k}}\right]^*\right)\bigg)
\end{split}
\end{equation}
with the three-index matrix element $\mathcal{M}^{\mu\alpha\beta}_{mn\mathbf{k}} = ir^\beta_{nm}r^{\alpha;\mu}_{mn}$ (identical to the ordinary shift current), and the four-index matrix element (suppressing $\mathbf{k}$-dependence of all variables)
\begin{equation}
\begin{split}
    \mathcal{M}^{\nu\mu\alpha\beta}_{mn} &= -\left(\frac{i}{2}Q^{\nu\beta}_{nm} + \frac{\varepsilon^{\nu\beta\rho}}{\omega_{mn}}M^\rho_{nm}\right)\left(i\nabla^\mu_\mathbf{k}+\mathcal{A}^\mu_m-\mathcal{A}^\mu_n\right)r^\alpha_{mn} \\
    &~~~~+r^\beta_{nm}\left(i\nabla^\mu_\mathbf{k}+\mathcal{A}^\mu_m-\mathcal{A}^\mu_n\right)\left(\frac{i}{2}Q^{\nu\alpha}_{mn}+\frac{\varepsilon^{\nu\alpha\rho}}{\omega_{mn}}M^\rho_{mn}\right) \\
    &~~~~+r^\beta_{nm}r^\alpha_{mn}\nabla^\nu_\mathbf{k}\left(\frac{\mathcal{A}^\mu_m+\mathcal{A}^\mu_n}{2}\right).
\end{split}
\end{equation}
In terms of the origin-independent multipole moments, we find (using the notation $\mathcal{O}^{;\mu}_{mn} \equiv (\nabla_\mathbf{k}-i(\mathcal{A}^\mu_m-\mathcal{A}^\mu_n))\mathcal{O}_{mn}$),
\begin{equation}
\begin{split}
    \mathcal{M}^{\nu\mu\alpha\beta}_{mn} &= \frac{1}{2}\widetilde{Q}^{\nu\beta}_{nm}r^{\alpha;\mu}_{mn} - \frac{1}{2}r^\beta_{nm}\widetilde{Q}^{\nu\alpha;\mu}_{mn} - i\frac{\varepsilon^{\nu\beta\rho}}{\omega_{nm}}\widetilde{M}^\rho_{nm}r^{\alpha;\mu}_{mn} + ir^\beta_{nm}\frac{\varepsilon^{\nu\alpha\rho}}{\omega_{mn}}\widetilde{M}^{\rho;\mu}_{mn} - i\frac{\varepsilon^{\nu\alpha\rho}}{(\omega_{mn})^2}(v^\mu_{m}+v^\mu_{n})r^\beta_{nm}\widetilde{M}^\rho_{mn} \\
    &~~~~+\frac{1}{2}\left(\mathcal{A}^\nu_m+\mathcal{A}^\nu_n\right)r^\beta_{nm}r^{\alpha;\mu}_{mn} - \frac{1}{2}r^\beta_{nm}\left(\mathcal{A}^\nu_m+\mathcal{A}^\nu_n\right)r^{\alpha;\mu}_{mn}\\
    &~~~~- \frac{1}{2}r^\beta_{nm}r^\alpha_{mn}\partial^\mu\left(\mathcal{A}^\nu_m+\mathcal{A}^\nu_n\right) + \frac{1}{2}r^\beta_{nm}r^\alpha_{mn}\partial^\nu(\mathcal{A}^\mu_m+\mathcal{A}^\mu_n)
\end{split}
\end{equation}
Remarkably, the origin-dependent terms from the multipole expansion (the parts that do not cancel out) can be combined with the last term to obtain a ``sum of Berry curvatures" term, where $\Omega^{\nu\mu}_{n} = \partial^\nu\mathcal{A}^\mu_n-\partial^\mu\mathcal{A}^\nu_n$ is the intraband Berry curvature. This term was also found in Ref. \cite{xie2025}, however, the origin-independent electric quadrupole term was missed. Finally, the full matrix element is obtained by adding the complex conjugate with $\alpha$ and $\beta$ interchanged.

For simplicity, and to emphasize the nontrivial geometry associated with the quadrupolar terms, the results in the main text are concerned with 1D models, for which all four spatial indices are equal. Dropping the spatial indices for simplicity, we obtain the simple form
\begin{align}
    \mathcal{M}^{(4),\text{1D}}_{mn} &= \frac{1}{2}\widetilde{Q}_{nm}\partial_k r_{mn}-\frac{1}{2}r_{nm}\partial_k\widetilde{Q}_{mn} + \frac{1}{2}\widetilde{Q}_{mn}\partial_k r_{nm}-\frac{1}{2}r_{mn}\partial_k\widetilde{Q}_{nm} -i\left(\mathcal{A}_m-\mathcal{A}_n\right)(r_{nm}\widetilde{Q}_{mn}-\widetilde{Q}_{nm}r_{mn}) \\
    &= \text{Re}\left(\widetilde{Q}_{nm}\partial_k r_{mn}-r_{nm}\partial_k\widetilde{Q}_{mn}\right) + 2(\mathcal{A}_m-\mathcal{A}_n)\text{Im}(\widetilde{Q}_{mn}r_{nm})
    .
\end{align}
Finally, the Fermi surface contribution $\propto f'_{n\mathbf{k}}$ vanishes for band insulators, which is the case considered in this work.

\section{For numerical calculations}

For numerical calculations, it is convenient to express quantities involving the Berry connection to quantities involving only the Bloch Hamiltonian, its $\mathbf{k}$-derivatives, and its eigenstates (with no derivatives acting on the eigenstates). For convenience, we summarize in this section the appropriate ``sum rules" for casting quantities in this form.

The interband Berry connection between two different bands $m$ and $n$ can be written as
\begin{equation}
    \mathcal{A}^\mu_{mn} = \frac{v^\mu_{mn}}{i\omega_{mn}}
\end{equation}
where $v^\mu_{mn} \equiv \frac{1}{\hbar}\langle u_{m\mathbf{k}} | \partial^\mu h_\mathbf{k} | u_{n\mathbf{k}} \rangle$ with $\partial^\mu\equiv \partial/\partial k_\mu$. This follows from $\partial^\mu\langle u_{m\mathbf{k}}|h_\mathbf{k}| u_{n\mathbf{k}} \rangle = 0$ for $m\ne n$, or equivalently from $v^\mu = i[H,r^\mu]$. In this manner, the origin-independent multipole moments can be expressed purely in terms of the interband velocity operator,
\begin{equation}
    \widetilde{Q}^{\nu\mu}_{mn} = \frac{1}{2}\sum_{l\ne m,n}\frac{v^\nu_{ml}v^\mu_{ln}}{\omega_{ml}\omega_{ln}} + [\mu\leftrightarrow\nu]
\end{equation}
\begin{equation}
    \epsilon^{\nu\mu\rho}\widetilde{M}^\rho_{mn} = \frac{1}{4}\sum_{l\ne m,n}\left(\frac{1}{\omega_{ml}}+\frac{1}{\omega_{ln}}\right)(v^\nu_{ml}v^\mu_{ln}+v^\mu_{ml}v^\nu_{ln}) + \frac{1}{2}\frac{v^\nu_{mn}}{\omega_{mn}}(v^\mu_{mm}+v^\mu_{nn}) -[\mu\leftrightarrow\nu]
\end{equation}
We remind the reader that the definition for the magnetic dipole moment differs from the definition in Ref. \cite{ocana2023} by the second term.

The shift current also depends on the ``generalized derivative" $\mathcal{O}^{;\mu}_{mn} \equiv (\partial^\mu-i(\mathcal{A}^\mu_{mm}-\mathcal{A}^\mu_{nn}))\mathcal{O}_{mn}$, where the intraband Berry connections are undesirable to compute directly numerically. Similar to above, we consider the quantity $\partial^\mu\partial^\nu \langle u_m|h_\mathbf{k}|u_n\rangle$ and find,
\begin{align}
    h^{\mu\alpha}_{mn} &= \partial^\mu v^\alpha_{mn} - \langle \partial^\mu u_{n\mathbf{k}} | v^\alpha_\mathbf{k} |u_{m\mathbf{k}}\rangle - \langle \partial^\mu u_{n\mathbf{k}} | v^\alpha_\mathbf{k} |u_{m\mathbf{k}}\rangle\\
    &= \partial^\mu v^\alpha_{mn} - i[\mathcal{A}^\mu,v^\alpha]_{mn} \\
    &= \partial^\mu v^\alpha_{mn} - i(\mathcal{A}^\mu_{mm}-\mathcal{A}^\mu_{nn})v^\alpha_{mn} -i\mathcal{A}^\mu_{mn}\left(v^\alpha_{mm}-v^\alpha_{nn}\right) - i\sum_{l\ne m,n}\left(\mathcal{A}^\mu_{ml}v^\alpha_{ln}-v^\alpha_{ml}\mathcal{A}^\mu_{ln}\right) \\
    &\equiv v^{\alpha;\mu}_{mn} -\frac{v^\mu_{mn}}{\omega_{mn}}\left(v^\alpha_{mm}-v^\alpha_{nn}\right) - \sum_{l\ne m,n}\left(\frac{v^\mu_{ml}v^\alpha_{ln}}{\omega_{ml}} -\frac{v^\alpha_{ml}v^\mu_{ln}}{\omega_{ln}}\right).
\end{align}
This implies that
\begin{align}
    r^{\alpha;\mu}_{mn} &\equiv \left(\frac{v^\alpha_{mn}}{i\omega_{mn}}\right)^{;\mu} \\
    &=-\partial^\mu\omega_{mn}\frac{v^\alpha_{mn}}{i(\omega_{mn})^2} + \frac{v^{\alpha;\mu}_{mn}}{i\omega_{mn}} \\
    &= \frac{h^{\mu\alpha}_{mn}}{i\omega_{mn}} + \frac{v^\mu_{mn}(v^\alpha_{mm}-v^\alpha_{nn})-(v^\mu_{mm}-v^\mu_{nn})v^\alpha_{mn}}{i(\omega_{mn})^2} + \sum_{l\ne m,n}\left(\frac{v^\mu_{ml}v^\alpha_{ln}}{\omega_{ml}} -\frac{v^\alpha_{ml}v^\mu_{ln}}{\omega_{ln}}\right)
\end{align}
where we used the fact that $\partial^\mu\omega_{mn} = v^\mu_{mm}-v^\mu_{nn}$.

We note that, alternatively, one can utilize the fully $\mathbf{q}$-dependent expression derived in the previous section and numerically compute it for very small $\mathbf{q}$, taking finite differences to numerically estimate the linear-in-$\mathbf{q}$ correction to the photocurrent. For convenience, we restate that result here,
\begin{align}
    \begin{split}
    \sigma^{\mu\alpha\beta}_\text{PV}(\omega,\mathbf{q}) = \frac{-\pi e^3}{\hbar^2\omega^2}\frac{1}{L^{2d}}\sum_{\mathbf{k},\mathbf{k}'}\delta_{\mathbf{k}'-\mathbf{k}-\mathbf{q}}\sum_{mn}f_{m\mathbf{k}',n\mathbf{k}}\,
        \delta(\omega_{m\mathbf{k}',n\mathbf{k}}-\omega)\left(\mathcal{M}^{\mu\alpha\beta}_{m\mathbf{k}',n\mathbf{k}}+\left[\mathcal{M}^{\mu\beta\alpha}_{m\mathbf{k}',n\mathbf{k}}\right]^*\right).
    \end{split}
\end{align}
\begin{equation}
    \mathcal{M}^{\mu\alpha\beta}_{m\mathbf{k}',n\mathbf{k}} = ij^\beta_{n\mathbf{k},m\mathbf{k}'}\left[j^{\mu\alpha}_{m\mathbf{k}',n\mathbf{k}}+\sum_l\left(\frac{j^\mu_{m\mathbf{k}',l\mathbf{k}'}j^\alpha_{l\mathbf{k}',n\mathbf{k}}}{\omega_{m\mathbf{k}',l\mathbf{k}'}+2i\eta}-\frac{j^\alpha_{m\mathbf{k}',l\mathbf{k}}j^\mu_{l\mathbf{k},n\mathbf{k}}}{\omega_{l\mathbf{k},n\mathbf{k}}+2i\eta}\right)\right]
\end{equation}
where $j^\mu_{m\mathbf{k}',n\mathbf{k}} \equiv \langle \psi_{m\mathbf{k}'}|j^\mu_{\mathbf{k}'-\mathbf{k}}|\psi_{n\mathbf{k}}\rangle$ denotes the finite-momentum matrix elements of the current operator, $\omega_{m\mathbf{k}',n\mathbf{k}} \equiv (\epsilon_{m\mathbf{k}'}-\epsilon_{n\mathbf{k}})/\hbar$ denotes band energy differences, $f_{m\mathbf{k}',n\mathbf{k}}\equiv f_\text{F-D}(\epsilon_{m\mathbf{k}'})-f_\text{F-D}(\epsilon_{n\mathbf{k}})$ denotes differences in Fermi factors, $L$ is the system size in each of the $d$ dimensions, and $\eta$ is a small phenomenological scattering rate.

\section{Scalar potential gauge}\label{scalar potential gauge}

One can alternatively model a spatially varying electric field (especially for one-dimensional systems) as a spatially varying scalar potential,
\begin{equation}
    V_{mn}(\omega,\mathbf{q}) = \frac{ie}{\hbar}\frac{E_\alpha(\omega)}{q_\alpha}\langle u_{m\mathbf{k}'}|\delta_{\mathbf{k}'-\mathbf{k}-\mathbf{q}} - \delta_{\mathbf{k}'-\mathbf{k}}e^{i\mathbf{q}\cdot\mathbf{r}_0}|u_{n\mathbf{k}}\rangle
\end{equation}
where $\mathbf{r}_0$ parametrizes an origin-dependent phase that we include for completeness, but will not affect final results. This potential is defined to recover the $\mathbf{q}\to 0$ limit $V_{\mathbf{q}\to 0} = e(\mathbf{r}-\mathbf{r}_0)\cdot\mathbf{E}(\omega)$, i.e. the ``length gauge" or ``dipole gauge" \cite{aversa1995}. Here, $\frac{E_\alpha(\omega)}{q_\alpha} \equiv \mathbf{E}(\omega)\cdot (\frac{\hat{\mathbf{q}}}{|\mathbf{q}|})$ denotes the component of the electric field along $\mathbf{q}$ divided by the magnitude of $\mathbf{q}$. This expression only includes the part of the light-matter coupling in the direction of the wave vector. Finally, only the parts proportional to $\delta_{\mathbf{k}'-\mathbf{k}-\mathbf{q}}$ survive, giving us
\begin{equation}
    \langle \vec{J}(0) \rangle^{(2)} = -\frac{e^3}{\hbar^2}\frac{E_\alpha(\omega)E_\beta(-\omega)}{q_\alpha q_\beta}\sum_{\mathbf{k}\mathbf{k}'}\delta_{\mathbf{k}'-\mathbf{k}-\mathbf{q}}\sum_{nm}\frac{f_{m\mathbf{k}',n\mathbf{k}}}{\omega+\omega_{m\mathbf{k}',n\mathbf{k}}+i\eta}\vec{\mathcal{M}}_{m\mathbf{k}',n\mathbf{k}} + [(\omega,\mathbf{q})\to (-\omega,-\mathbf{q})]
\end{equation}
\begin{equation}
    \vec{\mathcal{M}}_{m\mathbf{k}',n\mathbf{k}} = \langle u_{n\mathbf{k}}|u_{m\mathbf{k}'}\rangle\sum_{l}\left(\frac{\vec{v}_{m\mathbf{k}',l\mathbf{k}'}}{\omega_{l\mathbf{k}',m\mathbf{k}'}+2i\eta}\langle u_{l\mathbf{k}'}|u_{n\mathbf{k}}\rangle-\langle u_{m\mathbf{k}'}|u_{l\mathbf{k}}\rangle\frac{\vec{v}_{l\mathbf{k},n\mathbf{k}}}{\omega_{n\mathbf{k},l\mathbf{k}}+2i\eta}\right)
\end{equation}
The matrix element part $\vec{\mathcal{M}}_{m\mathbf{k}',n\mathbf{k}}$ is conventionally separated into ``band off-diagonal" and ``band on-diagonal" parts, which respectively give rise to the shift current and the injection current,
\begin{equation}
    \vec{\mathcal{M}}^\text{off-diag}_{m\mathbf{k}',n\mathbf{k}} = \langle u_{n\mathbf{k}}|u_{m\mathbf{k}'}\rangle\left(\sum_{l\neq m}\frac{\vec{v}_{m\mathbf{k}',l\mathbf{k}'}}{\omega_{l\mathbf{k}',m\mathbf{k}'}+2i\eta}\langle u_{l\mathbf{k}'}|u_{n\mathbf{k}}\rangle-\sum_{l\neq n}\langle u_{m\mathbf{k}'}|u_{l\mathbf{k}}\rangle\frac{\vec{v}_{l\mathbf{k},n\mathbf{k}}}{\omega_{n\mathbf{k},l\mathbf{k}}+2i\eta}\right)
\end{equation}
\begin{equation}
    \vec{\mathcal{M}}^\text{on-diag}_{m\mathbf{k}',n\mathbf{k}} = \frac{-i}{2\eta}|\langle u_{n\mathbf{k}}|u_{m\mathbf{k}'}\rangle|^2\left(\vec{v}_{m\mathbf{k}'}-\vec{v}_{n\mathbf{k}}\right)
\end{equation}
Assuming small $\eta$ compared to any energy difference $\omega_{ln}$ with $l\neq n$, we can $\eta$ to $0$ in the ``off-diag" expression and use $\vec{v}_{mn,\mathbf{k}} = i\omega_{mn}\langle u_{m\mathbf{k}}|\vec{\nabla}_\mathbf{k}u_{n\mathbf{k}}\rangle $ to recover a familiar form for the shift current,
\begin{align}
    \vec{\mathcal{M}}^\text{off-diag}_{m\mathbf{k}',n\mathbf{k}} &= \langle u_{n\mathbf{k}}|u_{m\mathbf{k}'}\rangle\left(-i\sum_{l\neq m}\langle u_{m\mathbf{k}'}|\vec{\nabla}_{\mathbf{k}'}u_{l\mathbf{k}'}\rangle\langle u_{l\mathbf{k}'}|u_{n\mathbf{k}}\rangle+i\sum_{l\neq n}\langle u_{m\mathbf{k}'}|u_{l\mathbf{k}}\rangle\langle u_{l\mathbf{k}}|\vec{\nabla}_{\mathbf{k}}u_{n\mathbf{k}}\rangle\right) \\
    &= \langle u_{n\mathbf{k}}|u_{m\mathbf{k}'}\rangle\left(i\sum_{l\neq m}\langle \vec{\nabla}_{\mathbf{k}'}u_{m\mathbf{k}'}|u_{l\mathbf{k}'}\rangle\langle u_{l\mathbf{k}'}|u_{n\mathbf{k}}\rangle+i\sum_{l\neq n}\langle u_{m\mathbf{k}'}|u_{l\mathbf{k}}\rangle\langle u_{l\mathbf{k}}|\vec{\nabla}_{\mathbf{k}}u_{n\mathbf{k}}\rangle\right) \\
    &= \langle u_{n\mathbf{k}}|u_{m\mathbf{k}'}\rangle\left(i\langle \vec{\nabla}_{\mathbf{k}'}u_{m\mathbf{k}'}|u_{n\mathbf{k}}\rangle+i\langle u_{m\mathbf{k}'}|\vec{\nabla}_{\mathbf{k}}u_{n\mathbf{k}}\rangle + (\vec{\mathcal{A}}_{m\mathbf{k}'}-\vec{\mathcal{A}}_{n\mathbf{k}})\langle u_{m\mathbf{k}'}|u_{n\mathbf{k}}\rangle\right) \\
    &= \langle u_{n\mathbf{k}}|u_{m\mathbf{k}'}\rangle\left(i\vec{\nabla}_{\mathbf{k}'}+i\vec{\nabla}_\mathbf{k} + \vec{\mathcal{A}}_{m\mathbf{k}'}-\vec{\mathcal{A}}_{n\mathbf{k}}\right)\langle u_{m\mathbf{k}'}|u_{n\mathbf{k}}\rangle
\end{align}
where in the second step we use $\vec{\nabla}_\mathbf{k}\langle u_{m\mathbf{k}}|u_{n\mathbf{k}}\rangle = 0$ and we define the intraband Berry connection $\vec{\mathcal{A}}_{n\mathbf{k}} = i\langle u_{n\mathbf{k}}|\vec{\nabla}_\mathbf{k}u_{n\mathbf{k}}\rangle$. In the $\mathbf{q}\to 0$ limit, this recovers the conventional shift current,
\begin{equation}
    \lim_{\mathbf{q}\to 0}\frac{E_\alpha(\omega)E_\beta(-\omega)}{q_\alpha q_\beta}\sum_{\mathbf{k}'}\delta_{\mathbf{k}'-\mathbf{k}-\mathbf{q}}\vec{\mathcal{M}}^\text{off-diag}_{m\mathbf{k}',n\mathbf{k}} = \hat{\mathbf{q}}\cdot\vec{\mathcal{A}}_{nm,\mathbf{k}}\left(i\vec{\nabla}_\mathbf{k}+\vec{\mathcal{A}}_{m\mathbf{k}}-\vec{\mathcal{A}}_{n\mathbf{k}}\right)\vec{\mathcal{A}}_{mn,\mathbf{k}}\cdot\hat{\mathbf{q}}
\end{equation}

\section{Analytical 3-band models with nontrivial quadrupolar response}

In the main text, we observed that because of the form of the interband quadrupole moment $Q^{\mu\nu}_{mn\mathbf{k}} = \frac{1}{2}\langle \check{\partial}^\mu u_{m\mathbf{k}} | \check{\partial}^\nu u_{n\mathbf{k}}\rangle + [\mu\leftrightarrow\nu]$ with $\check{\partial}^\mu = (1-|u_{m\mathbf{k}}\rangle\langle u_{m\mathbf{k}}|)(1-|u_{n\mathbf{k}}\rangle\langle u_{n\mathbf{k}}|)\partial/\partial k_\mu$, a model whose Hilbert space is only $\mathbb{CP}^1$ is insufficient for a quadrupolar optical response. This excludes, for instance, models of the form $h_\mathbf{k} = \mathbf{d}(\mathbf{k})\cdot\boldsymbol{\sigma}$ where $\left\{\sigma_i\right\}_{i=1,2,3}$ is any representation of the generators of the $SU(2)$ Lie algebra.

We are therefore motivated to consider subspaces of the next largest Hilbert space, $\mathbb{CP}^2$, that have simple analytical properties. We consider the following four matrices,
\begin{equation}
\begin{split}
\lambda_1 =
    \frac{1}{\sqrt{2}}
    \begin{pmatrix}
    0 & 1 & 0 \\
    1 & 0 & -1 \\
    0 & -1 & 0
    \end{pmatrix},~~~
\lambda_2 =
    \frac{1}{\sqrt{2}}
    \begin{pmatrix}
        0 & -i & 0 \\
        i & 0 & -i \\
        0 & i & 0
    \end{pmatrix},~~~
\lambda_3 =
    \begin{pmatrix}
    -1 & 0 & 0 \\
    0 & 0 & 0 \\
    0 & 0 & 1
    \end{pmatrix},~~~
    \lambda_4 =
    \begin{pmatrix}
    0 & 0 & -i \\
    0 & 0 & 0 \\
    i & 0 & 0
    \end{pmatrix}
    ,
\end{split}
\end{equation}
which generate a four-dimensional subspace of $SU(3)$. Bloch Hamiltonians of the form $\mathbf{d}(\mathbf{k})\cdot\boldsymbol{\lambda}$, where $\mathbf{d}(\mathbf{k}) \in \mathbb{R}^4$, have the nice property that the eigenvalues are $0, \pm d$ where $d=|\mathbf{d}(\mathbf{k})|$, and the eigenvectors can be written analytically as smooth functions of the unit vector $\hat{d}(\mathbf{k}) \equiv \mathbf{d}(\mathbf{k})/d$,
\begin{equation}
    h =
    \begin{pmatrix}
        -d_3 & \frac{1}{\sqrt{2}}\left(d_1 - i d_2\right) & -i d_4 \\
        \frac{1}{\sqrt{2}}\left(d_1 + i d_2\right) & 0 & \frac{1}{\sqrt{2}}\left(-d_1 - i d_2\right) \\
        id_4 & \frac{1}{\sqrt{2}}\left(-d_1 + i d_2\right) & d_3
    \end{pmatrix}
    ,
\end{equation}
\begin{equation}
    \mathcal{U}^\dagger h \mathcal{U} =
    \begin{pmatrix}
        -d & 0 & 0 \\
        0 & 0 & 0 \\
        0 & 0 & +d
    \end{pmatrix}
    ~~~\text{with}~~~
    \mathcal{U} =
    \begin{pmatrix}
        -\frac{1}{2}-\frac{d_3-id_4}{2d} & \frac{(d_1-id_2)(d_3-id_4)}{\sqrt{2}d\sqrt{d_3^2+d_4^2}} & \frac{1}{2}-\frac{d_3-id_4}{2d}\\
        \frac{d_1+id_2}{\sqrt{2}d}& \frac{\sqrt{d_3^2+d_4^2}}{d} & \frac{d_1+id_2}{\sqrt{2}d} \\
        \frac{1}{2}-\frac{d_3-id_4}{2d}& \frac{(d_1-id_2)(d_3-id_4)}{\sqrt{2}d\sqrt{d_3^2+d_4^2}} & -\frac{1}{2}-\frac{d_3-id_4}{2d}
    \end{pmatrix}
    .
\end{equation}
This is manifestly smooth as long as $d_3$ and $d_4$ are never simultaneously zero.

\section{Dimensional analysis of the continuum models of twisted bilayer MoTe$_2$}

We briefly comment on the order of magnitude of the linear-in-wavevector photovoltaic conductivity in twisted bilayer MoTe$_2$. We first note that we define the conductivity as a 2D conductivity: i.e. the current density has units of A/m. In units where the moir\'e lattice constant and $\frac{e^3}{\hbar^2}$ are taken to be unity and $\omega$ is measured in units of meV$/\hbar$, we numerically find values for $\sigma^{\nu\mu\alpha\beta}_{(1),\text{PV}}(\omega)$ on the order of unity. In SI units, we have that $\frac{e^3}{\hbar^2} = 3.697\times 10^{11}$ A V$^{-2}$ s$^{-1}$, the moir\'e lattice constant $a = 5.19\times 10^{-9}$ m, and $\hbar/[1\text{ meV}] = 6.582\times 10^{-13}$ s. In all, the relevant prefactor in front of the dimensionless numerical calculation is given by
\begin{align}
    \sigma^{\nu\mu\alpha\beta}_{(1),\text{phys.}}(\omega) &= \left(\frac{e^3}{\hbar^2}\times a^2\times \frac{\hbar}{1\text{ meV}}\right)\sigma^{\nu\mu\alpha\beta}_{(1),\text{num.}}(\omega) \notag \\
    &= (6.554\times 10^2 \text{A V$^{-2}$ nm$^2$})~\sigma^{\nu\mu\alpha\beta}_{(1),\text{num.}}(\omega).
\end{align}
Therefore, the results for the linear-in-wavevector photovoltaic conductivity in Figure 3 of the main text are on the order of $10^3$ A V$^{-2}$ nm$^2$. As a check, for a typical terahertz wavenumber of $q \sim 10^{-6}$ nm$^{-1}$, this means $q_\nu \sigma^{\nu\mu\alpha\beta}_{(1)}(\omega) \sim 10^3$ $\mu$A V$^{-2}$ nm, which compares well to previously reported calculations of wavevector-dependent photocurrents in twisted bilayer graphene \cite{lu2025}. Despite coming from centrosymmetric crystals, this magnitude of photovoltaic conductivity is comparable to the reported values of $10^2 - 10^4$ $\mu$A V$^{-2}$ nm in inversion-breaking 2D platforms \cite{lai2025,mao2025}.

\newpage

\section{Supplementary plots of the photocurrent in twisted bilayer MoTe$_2$}

\begin{figure}[h]
    \centering
    \includegraphics[width=0.75\linewidth]{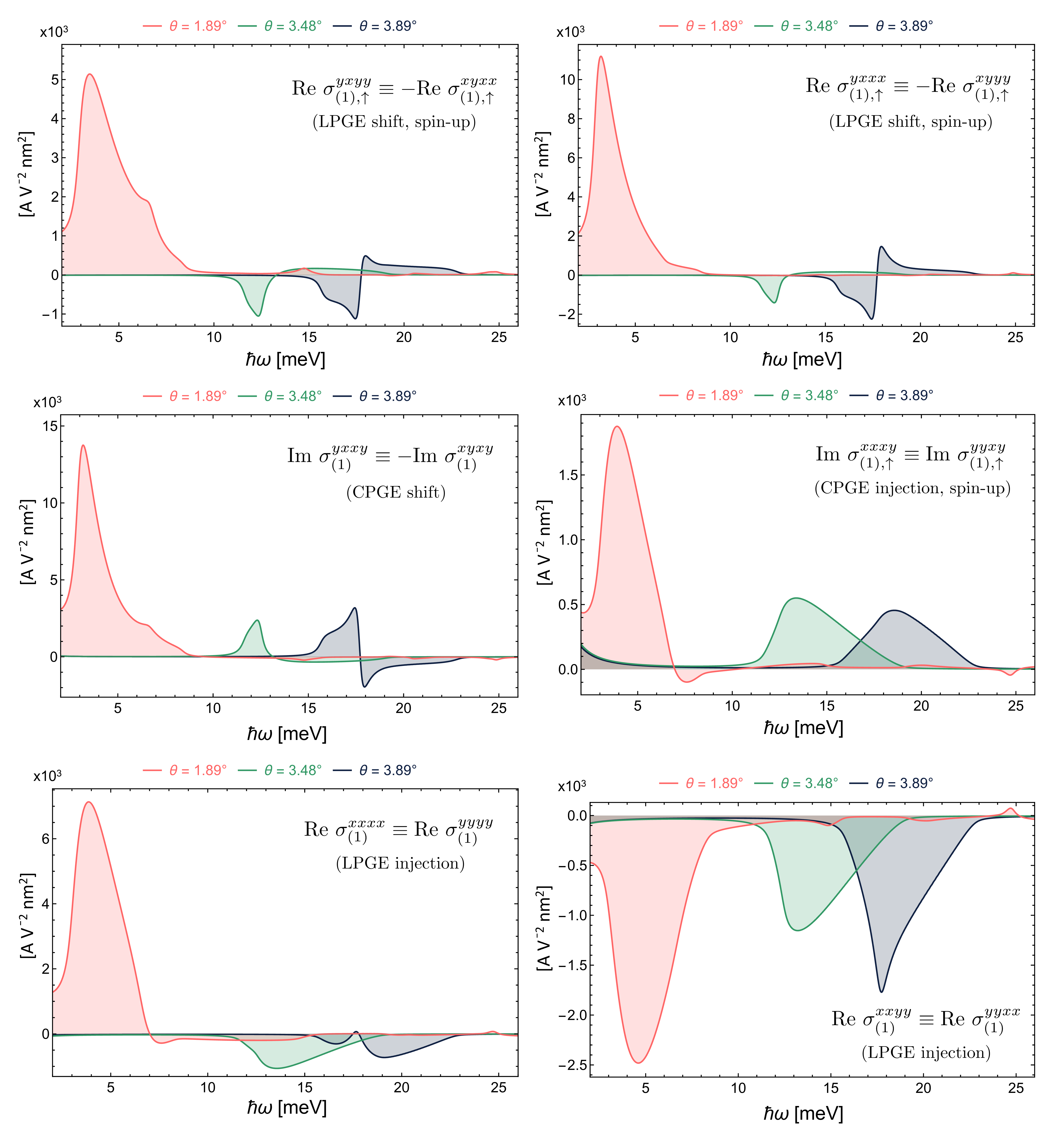}
    \caption{Full characterization of the linear-in-wavevector BVPE conductivity (here abbreviated as $\sigma^{\nu\mu\alpha\beta}_{(1)}$) for the continuum models of twisted MoTe$_2$ discussed in the main text. All other components of $\sigma_{(1)}^{\nu\mu\alpha\beta}$ are zero or related to one of the above components by symmetry (i.e. note that $\text{Im }\sigma^{\nu\mu\alpha\beta}_{(1)}$ is antisymmetric in $\alpha\leftrightarrow\beta$).}
    \label{fig:supplement CPGE and inj}
\end{figure}

\bibliography{references}